        \documentstyle[]{l-aa}
        \input psfig.tex

	\newcommand{\BV}{\mbox{${\rm (B-V)}$}}
	\newcommand{\VK}{\mbox{${\rm (V-K)}$}}
	\newcommand{\UB}{\mbox{${\rm (U-B)}$}}
	\newcommand{\UV}{\mbox{${\rm (U-V)}$}}
	
	\newcommand{\VR}{\mbox{${\rm (V-R)}$}}
	
	\newcommand{\UVex}{\mbox{${\rm (1550-V)}$}}
        \newcommand{\Hbeta}{\mbox{${\rm H_{\beta}}$}}
        \newcommand{\MgFe}{\mbox{${\rm [MgFe]} $}}
        \newcommand{\MFe}{\mbox{${\rm < Fe >} $}}
        
	\newcommand{\Mv}{\mbox{${\rm M_{V}}$}}
        \newcommand{\Mbol}{\mbox{${\rm M_{bol}}$}}
	
	\newcommand{\DMo}{\mbox{${\rm (m-M)_{0}}$}}
	\newcommand{\FeH}{\mbox{{\rm [Fe/H]}}}
	\newcommand{\Msun}{\mbox{${\rm M_{\odot}}$}}
	\newcommand{\Teff}{\mbox{${\rm T\sub{eff}}$}}
        
        \newcommand{\dydz}{\mbox{${\rm \Delta Y/ \Delta Z}$}}
	
       	\newcommand{\logS}{\mbox{${\rm \log\Sigma }$}}
   	
        \newcommand{\MS}{\mbox{${\rm M_{S}}$}}
        \newcommand{\ML}{\mbox{${\rm M_{L}}$}}
        \newcommand{\MD}{\mbox{${\rm M_{D}}$}}
        \newcommand{\RL}{\mbox{${\rm R_{L}}$}}
        \newcommand{\RD}{\mbox{${\rm R_{D}}$}}

	\newcommand{\sub}[1]{\mbox{$_{\rm #1}$}}
	
        \def\oneskip{\vskip 8pt}	
        \def\smallskip{\vskip 6pt}
        \def\littleskip{\vskip 4pt}
        \def\M12{${\rm M_{12}}$}

\begin{document}

	\thesaurus{    }

        \title{Probing the Age of Elliptical Galaxies  }

	\author{  A. Bressan$^1$,  C. Chiosi$^2$, and R. Tantalo$^2$  }
	\institute{
                  $^1$ Astronomical Observatory, Vicolo dell'Osservatorio 5, 
                   35122 Padova, Italy\\
                  $^2$ Department of Astronomy, University of Padova,
      Vicolo dell'Osservatorio 5, 35122 Padova, Italy \\         }

	\offprints{C. Chiosi  }

	\date{Received April  1995; accepted }

	\maketitle
	\markboth{Ages of Elliptical Galaxies }{}

\begin{abstract}
In this paper we address the question whether age and metallicity effects can
be disentangled with the aid of the  broad-band colours and spectral indices
from  absorption feature strengths, so that the age of elliptical galaxies can
be inferred. The observational data under examination are the  indices \Hbeta\
and \MgFe,  and the velocity dispersion $\Sigma$ for the sample of galaxies of
Gonzales (1993), supplemented by the ultra-violet data, i.e. the colour
(1550-V), of Burstein et al. (1988). The analysis is performed with the aid of
chemo-spectro-photometric models of elliptical galaxies with infall of
primordial gas (aimed at simulating the collapse phase of galaxy formation) and
the occurrence of galactic winds. The galaxy models are from Tantalo et al.
(1995). The study consists of four  parts. In the first one, the aims are
outlined and the key data are presented. In the second part, we summarize the
main properties of the infall models that are relevant to our purposes. In the
third part we present the  detailed calculations of the spectral indices for
single stellar populations and model galaxies. To this aim, we use the
analytical relations of Worthey et al. (1994) who  give index strengths as a
function of stellar parameters. In the last part, we examine the
age-metallicity problem. In contrast with previous interpretations of the
\Hbeta\ and \MgFe\ data  as a sort of age sequence (Gonzales 1993), we find
that the situation is more complicate when the space of the four variables
\Hbeta,  \MgFe, (1550-V), and $\Sigma$ is examined. Galaxies in the \Hbeta\ and
\MgFe\ plane do not follow a pure sequence either of age or metallicity. The
observed (1550-V) colours are not compatible with young ages. Basically, all
the galaxies in the sample are old objects (say as old as 13$\div$15 Gyr) but
have suffered from  different histories  of star formation. Specifically,  it
seems that some galaxies  have exhausted the star forming activity at very
early epochs with no significant later episodes. Others have continued to form
stars for long periods of time. This is perhaps sustained by the analysis of
the gradients in the \Hbeta\ and \MgFe\ indices across the galaxies. There are
galaxies with no age difference among the various regions.  There are other
galaxies in  which large gradients in the mean age of the star forming activity
between the central and the peripheral regions seem to exist. The nucleus turns
out to be younger and more metal-rich than the outer regions. Finally, there
are galaxies in which the nucleus is older but less metal-rich than the
external regions. All this perhaps hints not only different histories of star
formation  but also different mechanisms of galaxy formation difficult to pin
down at the present time. From the analysis of the  \Hbeta,  \MgFe, (1550-V),
and $\Sigma$ space, and of the age and metallicity gradients in single
galaxies, the suggestion is advanced that the overall duration of the star
forming activity is inversely proportional to the velocity dispersion $\Sigma$
(and perhaps  galactic mass). 


\end{abstract}

\section {Introduction}

These days, it is understood that age and metallicity govern the properties of
the stellar populations in galaxies of different morphological type.
Unfortunately, even in the simplest case of early-type galaxies (ellipticals),
age effects  often mask metallicity effects so that separating the two is a
cumbersome affair (cf. Renzini 1986; Buzzoni et al. 1992, 1993; Worthey 1994 
for more details). The age-metallicity dilemma originates from the fact that 
increasing either the metallicity or the age makes the integrated spectral
energy distribution (ISED) of a single stellar population (SSP) redder. 

The subject of galaxy ages  has been tackled with different techniques of
populations synthesis (evolutionary and optimizing,  broad-band colours and
narrow-band indices) with rather contrasting results. 

From evolutionary synthesis models  and broad-band colours  the indication
arises that elliptical galaxies are made of old stellar populations moderately
metal rich and age of about 15 Gyr. Supplementary episodes of star formation at
more recent epochs seem to be excluded. 

In contrast,  the optimized synthesis models (cf. O' Connell 1986 and
references) suggest  that large age spreads are possible with a substantial
contribution from  young stars  to the integrated light (for the prototype
galaxy M32 as young as 5 Gyr). 
 
Recently, the study of spectral indices reveals a powerful tool to discriminate
age from metallicity effects. According to Buzzoni et al. (1992, 1993) the use
of calibrated  Mg2 and \Hbeta\ indices  strengthens  the conclusion reached
with the broad-band colours. On the contrary, according to Gonzales (1993 and
references) the indices \MgFe\ and  \Hbeta\ hint that a substantial age spread
ought to exist. 

The problem is further complicated by the poor knowledge of the star formation
history of elliptical galaxies. Indeed the view that all elliptical galaxies
are primordial old objects (cf. Bower et al. 1992a,b) coexists with the view
that every elliptical galaxy is the product of merger events involving smaller
galaxies (cf. Schweizer \& Seitzer 1992, Alfensleben \& Gerhard 1994, Charlot
\& Silk 1994). 

In this work we would like to 
try a more articulated analysis aimed at separating age
from metallicity effects. The study is  based on the recent models for
elliptical galaxies  by Bressan et al. (1994), Tantalo (1994),  and Tantalo et
al. (1995), however implemented with the calibrations for  narrow band indices
obtained by Worthey et al. (1994). In order to separate the 
age from metallicity we
look at the properties of the galaxies in the  four dimensional space of the
parameters \Hbeta, \MgFe, \UVex, and  velocity dispersion $\Sigma$.  The
indices \Hbeta\ and \MgFe\ are particularly suited to separate age from
metallicity effects because \Hbeta\ is a measure of the turn-off colour and
luminosity,  and age in turn, whereas  \MgFe\ is more sensitive to the RGB
colour and hence metallicity (Buzzoni et al. 1992, 1993, Gonzales 1993). 

The paper is organized in the following way. Section 2  briefly summarizes
current information on basic data for elliptical galaxies, i.e. the
observational hints on metallicities and age ranges for the stellar content of
elliptical galaxies and bulges, the line strength indices \Hbeta\ and \MgFe\,
the colour-magnitude relation (CMR) of elliptical galaxies and its current
interpretation. Section 3 describes the recent models for  elliptical galaxies
by Bressan et al. (1994) and Tantalo et al. (1995). 
 Section 4 presents the calculation of the   narrow
band indices for SSPs. Section 5 deals with  the integrated narrow band indices
for models of elliptical galaxies, with particular attention to the evolution
in the \Hbeta\ - \MgFe\ plane. Section 6 addresses the question whether the
observed distribution in the \Hbeta\ - \MgFe\ plane could result from recent
bursts of star formation superposed to much older populations. Section 7
analyses data and theoretical predictions in the four dimensional space of
coordinates \Hbeta, \MgFe, \UVex, and $\Sigma$, and presents a general
discussion of the results emerging from this study. Section 8 deals with the
gradients in \Hbeta\ and \MgFe\ across individual galaxies, and proposes a new
method to separate age from metallicity. In particular it shows that different
regions of a galaxy may have different age and metallicity and provides clues
to understand the mechanism of galaxy formation.   Finally, Section 9 presents
some concluding remarks.

\section{Basic Observational data }

In this section we summarize current information on basic data for elliptical
galaxies that are either constraints or targets of our models. 

\subsection{ Hints on metallicities and ages }

Direct determinations of metallicities in elliptical galaxies are still rather
limited. Most of the information comes from metal line-strength indices and
their radial variations (Carollo et al. 1993, Carollo \& Danziger 1994,  Davies
et al. 1993)  or colour gradients (Schombert et al. 1993). The evidence  arises
that the metallicities for elliptical galaxies are solar or larger.
Furthermore, for bright elliptical galaxies there are also indications that the
$\alpha$-elements (O, Mg, Si, etc..) are enhanced with respect to Fe. The
average [Mg/Fe], in particular,  exceeds that of the most metal-rich stars in
the solar vicinity by about 0.2-0.3 dex, and the ratio [Mg/Fe] is expected to
increase with the galactic mass up to the this value. Due to their proximity,
the metallicity  of stars in the Galactic Bulge  is currently measured by
spectroscopic methods (cf. Rich 1990,  McWilliam \& Rich 1994). The mean
metallicity is slightly lower than solar and long tails towards the metal-poor
(${\rm [Fe/H]=-1}$) and metal-rich end (${\rm [Fe/H]=1}$) are also present.
High metallicities seem also to be indicated by the color-magnitude diagrams
(CMD) of stellar populations  in the bulges of nearby Galaxies (M31 and M32).
In the case of M32 (Freedman 1989, 1992), these stars are most likely in the
RGB and AGB phases and metallicities as high as [M/H]=0.1 are possible (cf. 
Bica et al. 1990, 1991). Concerning the age, from the magnitude of the
brightest AGB stars in M32, Freedman (1989, 1992) concludes that  a fraction of
these could be as young as 5 Gyr.

\begin{figure}
\psfig{file=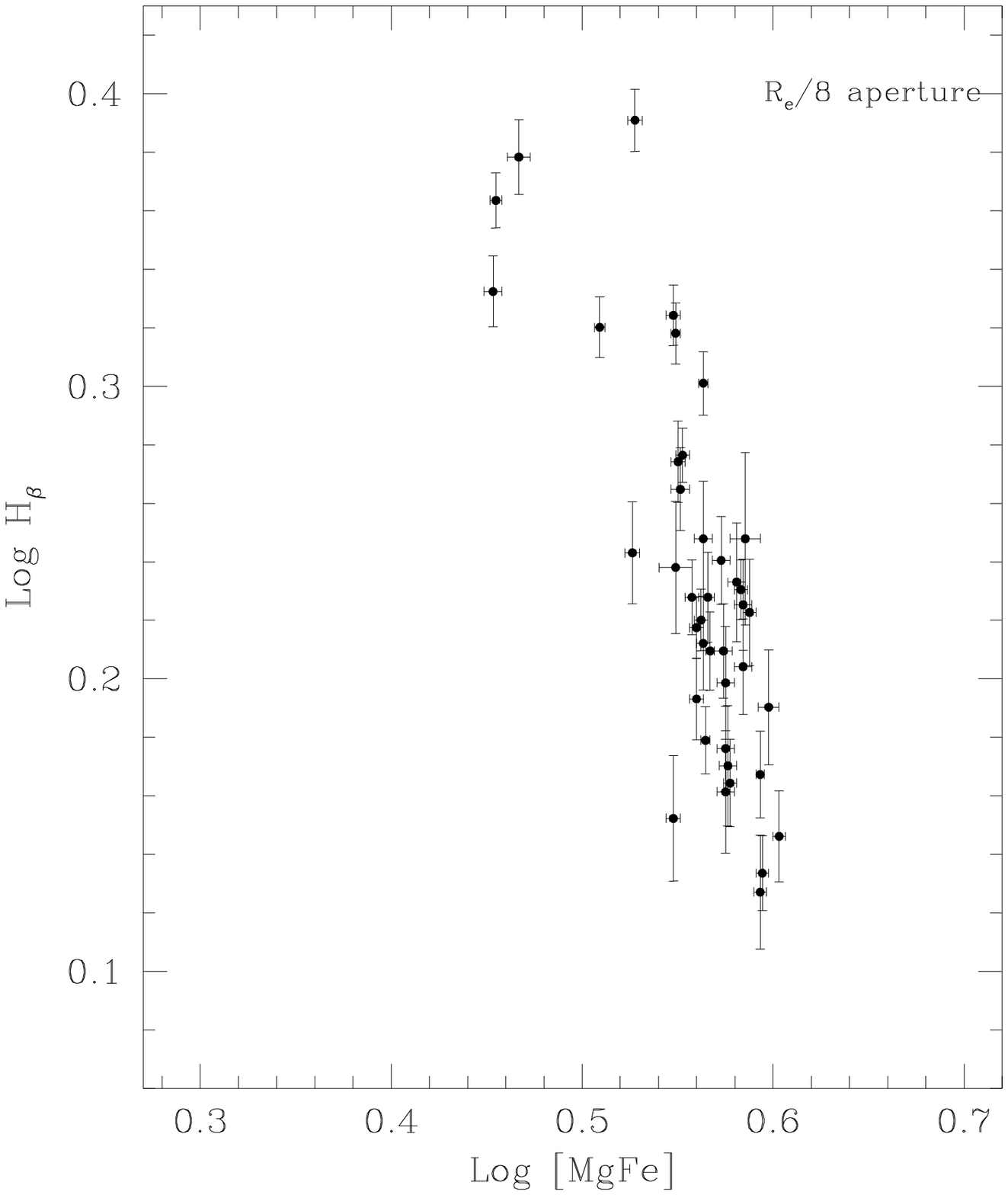,height=8.5truecm,width=8.5truecm}
\caption {The fundamental plane \Hbeta\ versus \MgFe\ for the Re/8 
data of Gonzales (1993) together with the error bars for each
galaxy } 
\label{gonz_data}
\end{figure}

\subsection {Line strength indices \Hbeta\ and \MgFe\ }

Table~\ref{tab_data} contains  the key data relevant to our purposes, namely
the fully corrected indices \Hbeta\ and \MgFe\, and the velocity dispersion
$\Sigma$ (km sec$^{-1}$) of Gonzales (1993) for two different galaxy coverage,
i.e. the  central region within Re/8, where Re is the effective radius, and the
wider region within Re/2. Column (1) is the galaxy identification NGC number,
column (2) the logarithm of the \Hbeta\ index, column (3) is the logarithm of
the \MgFe\ index, column (4) is the logarithm of the velocity dispersion 
$\Sigma$ in km sec$^{-1}$ for the Re/8-data. Columns (5), (6) and (7) show the
same but for the Re/2-data. The content of the remaining columns will be
described below.  

The fundamental plane \Hbeta\ versus \MgFe\ for the Re/8 
data is shown in Fig.~\ref{gonz_data} together with the error bars for each
galaxy. 
It is worth noticing that passing from the Re/8 to Re/2 data, there is a mean
decrease in $\log\MgFe$ amounting 0.032,  and a mean increase in $\log\Hbeta$
amounting to 0.003.

\begin{table*}
\caption{Basic observational data used in this study: fully corrected 
indices \Hbeta\ and \MgFe, and velocity dispersion 
$\Sigma$ (km sec$^{-1}$) for
the Re/8 and Re/2 data of Gonzales (1993);  \UVex\  colours of 
Burstein et al. (1988) for the central regions; magnitudes ${\rm M_V}$ and
${\rm M_V}$,  and colours (B--V) and (V--K) of Bender et al. (1992, 
1993), Bower et al. (1992), Marsiaj (1992) and Schweizer \& Seitzer (1992) 
as explained in the text. } 
\littleskip
\scriptsize
\begin{center}
\begin{tabular*}{180mm}{ccc ccc ccc ccc ccc cc} 
\hline 
\hline
NGC &\Hbeta&\MgFe&$\Sigma$&\Hbeta&\MgFe&$\Sigma$&\UVex&Note&-${\rm M_B}$&S&
-${\rm M_B}$& B-V& B-V& V-K& V-K&-$M_V$\\
\hline
(1) &(2) &(3) &(4) &(5) &(6)&(7)  &(8)& (9) &(10)&(11)&(12)&(13)&(14)&
(15)&(16)&(17)\\
\hline
    &Re/8&Re/8&Re/8&R2/2&Re/2&Re/2&    &   &     &   &     &    &    &    &    &     \\
\hline  
 221&0.36&0.45&1.86&0.33&0.45&1.81&4.50&   &     &   &15.70&0.84&    &    &    &16.64\\
 224&0.22&0.59&2.19&0.23&0.58&2.19&3.51&   &     &   &     &    &0.87&    &    &     \\
 315&0.24&0.57&2.51&0.25&0.54&2.49&    &   &     &   &23.61&1.00&    &    &    &24.61\\
 507&0.24&0.55&2.42&0.31&0.53&2.42&    &   &     &   &     &    &    &    &    &     \\
 547&0.20&0.58&2.37&0.15&0.55&2.34&    &   &     &   &     &    &    &    &    &     \\
 584&0.32&0.55&2.29&0.31&0.52&2.23&3.93&   &     &   &21.72&0.93&0.94&    &3.18&22.66\\
 636&0.28&0.55&2.20&0.27&0.52&2.17&    &   &     &   &20.65&0.92&0.92&    &    &21.57\\  
 720&0.25&0.59&2.38&0.36&0.58&2.25&    &   &     &   &21.60&0.99&0.97&    &    &22.59\\
 821&0.22&0.56&2.28&0.26&0.53&2.25&    &   &     &   &21.61&0.98&0.98&    &    &22.59\\
1453&0.20&0.58&2.46&0.23&0.55&2.42&    & SS&22.26&1.5&     &    &1.02&    &    &     \\
1600&0.19&0.60&2.50&0.24&0.60&2.49&    &   &     &   &23.17&0.97&0.98&    &3.34&24.14\\
1700&0.32&0.55&2.36&0.32&0.52&2.35&    & SS&22.50&3.7&22.28&0.92&0.91&    &3.18&23.20\\
2300&0.23&0.58&2.40&0.21&0.56&2.35&    & SS&21.45&2.8&21.56&1.04&1.02&    &3.42&22.60\\
2778&0.25&0.56&2.19&0.19&0.53&2.11&    &   &     &   &18.14&0.91&    &    &    &19.05\\
3377&0.32&0.51&2.03&0.33&0.45&1.95&    & SS&19.06&1.5&19.49&0.90&0.90&    &2.95&20.39\\
3379&0.21&0.57&2.31&0.20&0.54&2.27&3.86& SS&20.44&0. &20.17&0.97&0.96&    &3.26&21.13\\
3608&0.23&0.57&2.25&0.24&0.53&2.21&    & SS&19.87&0. &19.87&0.97&0.94&    &3.17&20.84\\
3818&0.23&0.58&2.24&0.26&0.52&2.18&    & SS&19.61&1.3&19.49&0.92&0.92&    &    &20.41\\
4261&0.13&0.59&2.46&0.11&0.56&2.43&    & SS&21.78&1.0&21.74&0.99&0.99&    &3.29&22.73\\
4278&0.19&0.56&2.37&0.22&0.53&2.31&2.88& SS&19.93&1.5&19.79&0.96&    &    &3.21&20.75\\
4374&0.18&0.56&2.45&0.19&0.54&2.43&3.55&SSV&21.80&2.3&21.85&0.98&0.95&3.28&3.29&22.73\\
4472&0.21&0.57&2.45&0.22&0.56&2.41&3.42& V &     &   &22.21&0.98&0.96&3.32&3.36&23.10\\    
4478&0.26&0.55&2.11&0.24&0.52&2.13&    & V &     &   &19.44&0.89&    &3.13&3.24&20.33\\
4489&0.38&0.47&1.67&0.36&0.42&1.61&    &   &     &   &     &    &    &    &    &     \\
4552&0.17&0.59&2.40&0.18&0.56&2.37&2.35& V &     &   &     &    &0.96&3.33&3.31&21.68\\
4649&0.15&0.60&2.49&0.14&0.57&2.45&2.24&   &     &   &21.81&1.01&0.99&    &3.36&22.82\\
4697&0.24&0.53&2.21&0.22&0.47&2.19&3.41&SSV&21.76&0. &21.51&0.95&0.94&    &3.22&22.46\\
5638&0.22&0.56&2.19&0.23&0.52&2.14&    &   &     &   &     &    &    &    &    &     \\
5812&0.23&0.58&2.30&0.23&0.56&2.26&    &   &     &   &     &    &    &    &    &     \\
5813&0.15&0.55&2.31&0.09&0.53&2.32&    &   &     &   &     &    &    &    &3.29&     \\
5831&0.30&0.56&2.20&0.31&0.57&2.18&    & SS&20.28&2.6&20.22&0.95&    &    &3.42&21.85\\
5846&0.16&0.58&2.35&0.10&0.54&2.32&3.11&   &     &   &21.85&0.99&    &    &3.42&22.84\\
6127&0.18&0.58&2.38&0.18&0.55&2.34&    &   &     &   &     &    &    &    &    &     \\
6702&0.39&0.53&2.24&0.40&0.51&2.22&    &   &     &   &     &    &    &    &3.26&     \\
6703&0.27&0.55&2.26&0.26&0.51&2.22&    &   &     &   &     &    &    &    &    &     \\
7052&0.17&0.58&2.44&0.25&0.55&2.42&    &   &     &   &     &    &    &    &    &     \\
7454&0.33&0.45&2.03&0.32&0.40&2.04&    &   &     &   &     &    &    &    &    &     \\
7562&0.23&0.51&2.39&0.24&0.54&2.38&    &   &     &   &     &    &    &    &3.30&     \\
7619&0.13&0.59&2.48&0.17&0.56&2.44&    & SS&22.23&0. &22.35&1.01&1.02&    &3.47&23.36\\
7626&0.16&0.58&2.40&0.16&0.54&2.38&    & SS&22.28&2.6&22.36&1.00&1.00&    &3.39&23.36\\
7785&0.21&0.56&2.38&0.18&0.54&2.34&    &   &     &   &22.22&0.95&    &    &    &     \\
\hline 
\hline
\end{tabular*}
\end{center}
\label{tab_data}
\normalsize
\end{table*}

\subsection{  UV excess and  \UVex\ colour }

Much of the available information is from the study of Burstein et al. (1988).
The main points are the following: 
\littleskip
\noindent
(1) All studied elliptical galaxies have detectable flux short-ward of $\sim
2000$ \AA, with  large galaxy to galaxy differences in the level of the UV
flux.  The intensity of the UV emission is measured by the colour (1550-V). 
\littleskip
\noindent
(2) The colour (1550-V) correlates with the index Mg2, the velocity dispersion
$\Sigma$ and the luminosity (mass) of the galaxy. The few galaxies (e.g. 
NGC~205) in which active star  formation is seen do not obey these relations. 
\littleskip
\noindent
(3) An important constraint is posed by the HUT observations by Ferguson et al.
(1991) and Ferguson \& Davidsen (1993)  of the UV excess in the bulge of M31.
In this case the UV emission  shows a drop-off short-ward of about 1000 \AA\
whose interpretation requires that  the temperature of the emitting source must
be about 25,000 K.  Only a small percentage of the $912 \leq \lambda \leq 1200
$ \AA\ flux can be coming from stars hotter than 30,000 K and cooler than
20,000 K. 
\littleskip

The \UVex\ colours of Burstein et al. (1988) are listed in column (8) of
Table~\ref{tab_data}.

\subsection{The CMR  of early-type galaxies}

Colours and magnitudes of elliptical galaxies define a mean CMR, according to
which the colours get redder at decreasing absolute magnitude and hence
increasing mass of the galaxies. 

The slope of the CMR is commonly understood as indicative of a 
mass-metallicity sequence (Dressler 1984, Vader 1986) resulting from the effect
of galactic winds  (Larson 1974; Saito 1979a,b; Bressan et al. 1994). In brief,
independently of their total mass,  galaxies evolving as either {\it closed or
open boxes} (cf. Tinsley 1980) should reach metallicities governed by the
degree of gas gas consumption. Elliptical galaxies, having exhausted their
gas,  should possess almost identical (high) metallicities. 
Therefore, in order to have a mass-metallicity
sequence, a   mechanism  halting star formation and thus leaving different
metallicities must be invoked: galactic winds are the right agent (cf.  Arimoto
\& Yoshii 1986, 1987; Matteucci \& Tornamb\'e 1987; Yoshii \& Arimoto 1987;
Bressan et al. 1994; Tantalo et al. 1995). 
                                                                            
The tightness of the CMR seems to depend on the galaxy environment and could
reflect the process of galaxy formation. 

Looking at the CMR of Bower et al. (1992a,b) for galaxies in the largest known
nearby clusters Virgo and Coma, the colour dispersion is very small
(uncertainty in absolute magnitudes is less of a problem here), with typical
rms scatter of  0.05 mag of which 0.03 mag can be accounted for by
observational errors. Bower et al. (1992a,b) using the rate of colour change
${\partial \UV / \partial t}$ for SSPs of Bruzual (1983) argued that for the
Hubble time $t_H=15$ Gyr and coordination in the galaxy formation process
(their parameter $\beta=1$)  the galaxy ages are confined in the
range $13.5 \simeq t_F \simeq 15.0 $ Gyr. It is an easy matter to check that
the Bower et al. (1992a,b) conclusion neither depends on the particular
rate of colour change they have adopted nor the use of single SSPs instead of a
manyfold of SSPs with different chemical composition weighted on the past rate
of star formation (thus closer to the complexity of a real galaxy). This can be
checked using the SSPs of Bressan et al. (1994) and Tantalo et al. (1995) or
the models presented below. 
 
In contrast,  examining the colour dispersion (relatively  small also in this
case) of the CMR for nearby galaxies in small groups and field, Schweizer \&
Seitzer (1992) found it compatible with recent mergers of spiral galaxies
spread over several billion years. Also in this case the CMR implies a
mass-metallicity sequence. 

Since 
the  interpretation of the CMR bears very much on the kind of process of galaxy
formation and evolution, its use as a 
constrain on the galactic models has different implications.
 If the CMR is indicating nearly coeval, old galaxies
ranked along a mass-metallicity sequence,  passive evolution after the initial
star forming activity and onset of galactic winds is appropriate. This
facilitates the use of the CMR as a constrain on galactic models (cf. Bressan
et al. 1994, Tantalo et al. 1995). In contrast, if the CMR is compatible with 
mergers spreading over long periods of time, first the evolutionary history of
an elliptical galaxy is more complicated, and the use of the CMR as a
constraint is less secure. 

In order to cast light on this topic we try a closer inspection of the data and
derive the CMR for the Gonzales (1993) sample to be compared with that of Bower
et al. (1992a,b) and Schweizer \& Seitzer (1992). Cross-checking of the three
lists reveals that,  while the Gonzales (1993) and the Schweizer \& Seitzer
(1992) samples have many objects in common (17 galaxies), there is only one
galaxy in common to all the three groups (NGC~4374),  and four in common to
Gonzales (1993) and Bower et al. (1992a,b), namely the  galaxy above plus 
NGC~4472, NGC~4478, and NGC~4552. The galaxy  NGC~4374 is particularly interesting
because Schweizer \& Seitzer (1992) have estimated for it a merger age of about
7 Gyr. 

In addition to this, we tried to assign magnitudes and colours to as many 
 of the Gonzales (1993) galaxies as
possible making use  of magnitudes and colours
taken from Bender et al. (1992, 1993) and  Marsiaj (1992). All the data are
presented in Table~\ref{tab_data}. Column (9) indicates whether the galaxy
belong to the Schweizer \& Seitzer (1992) list (SS) and/or the Virgo cluster (V
or SSV); columns (10) and (11) are the ${\rm M_B}$ magnitude and the fine
structure parameter S, respectively,  of Schweizer \& Seitzer (1992). The
parameter S is a rough measure of dynamical youth or rejuvenation (0 no trace
of fine structure, 10 strong fine structure). Column (12) and (13) are the
magnitude ${\rm M_B}$ and colour (B-V), respectively, taken from Bender et al.
(1992, 1993). In a few cases these magnitudes are somewhat different from those
of Schweizer \& Seitzer (1992). Columns (14) is the (B-V) colour listed by
Marsiaj (1992). In general, it agrees with that of Bender et al. (1992, 1993).
Columns (15) and (16) are the (V-K) colours of the Bower et al. (1992a,b) and
Marsiaj (1992) lists, respectively. Also in this case,  for the few galaxies in
common  the agreement is remarkable. This means that Marsiaj's (1992) (V-K)
colours assigned to the remaining galaxies (when possible) can be considered as
reasonably good. Finally, column (17) lists the magnitude ${\rm M_V}$ obtained
from columns (12) and (13). 

\begin{figure}
\psfig{file=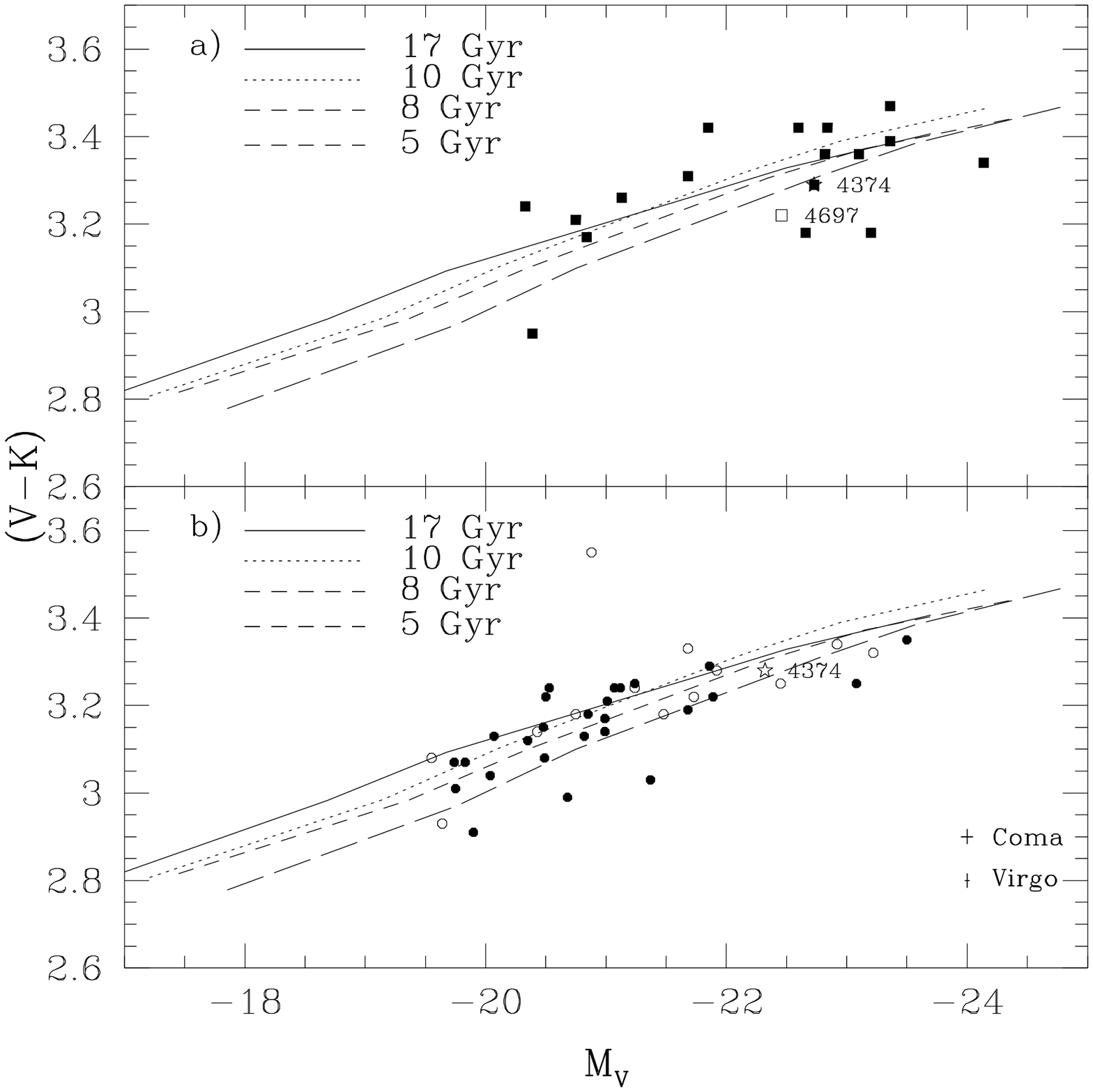,height=9.0truecm,width=8.5truecm}
\caption { {\bf Panel a}: the CMR for the sample by Schweizer \& Seitzer for
which we have obtained the magnitude $M_{V}$. The position of NGC~4374 
(filled star) and NGC~4697 (open square) is indicated.
 {\bf Panel b}: the CMR for
galaxies in Virgo (circles) and Coma (filled circles)
reproduced from Bower et al.(1992).  The galaxy NGC~4374 is shown by the open 
star. The typical error bars of the data 
are also displayed.  In addition to this, in each panel, are 
drawn the line of constant age for  the infall models  
characterized by the parameters:  $\tau=0.10$ Gyr, $k=1$, $\zeta=0.50$, 
and  $\nu$
variable with \ML\ (cf. Table 2). Each line corresponds to different values 
of the age: (solid: 17 Gyr), (dotted: 10 Gyr), (short-dashed: 8 Gyr), and 
(long-dashed: 5 Gyr). See the text for more details  } 
\label{cmr}
\end{figure}

The resulting \VK\ versus ${\rm M_V}$ CMR is shown in panel (a) of
Fig.~\ref{cmr}, whereas that of Bower et al. (1992a,b) is displayed in panel
(b). In this latter, we adopt for Virgo the distance modulus \DMo=31.54 (Branch
\& Tammann 1992) and apply to Coma the shift $\delta\DMo=3.65$ (cf. Bower et
al. 1992a,b). In Fig.~\ref{cmr} we also plot the loci of constant ages (17, 10,
8 and 5 Gyr) for the models presented in Sect. 3. The comparison of the two
CMRs yields the following results: 
\littleskip
\noindent
(1) The slopes of the two CMRs agree each other and with the theoretical
prediction. This means that both groups of galaxies reflect a similar
mass-metallicity sequence. 
\littleskip
\noindent
(2) Indeed the  Bower et al. (1992a,b) CMR is less scattered than the one of 
Schweizer \& Seitzer (1992), thus explaining the different interpretations. 
\littleskip
\noindent
3) NGC~4374 and NGC~4697, for which  rather recent merger and accompanying
stellar activities have  been proposed by Schweizer \& Seitzer (1992), are
compatible with the Bower et al. (1992a,b) conclusion at the same time. 
\littleskip
\noindent
4) Finally, as galactic models are imposed to match the slope and the mean 
(V--K) colours of the CMR, the results do not depend on the particular CMR 
in usage.  Therefore, owing to its better definition,
we prefer to adopt  the Bower et al. (1992)
CMR to infer the slope of the underlying mass-metallicity sequence.

\section{ Theoretical models of elliptical galaxies } 
In this section we summarize  the main properties of the models of elliptical
galaxies used in the present analysis. They are taken from the studies of
Bressan et al. (1994), Tantalo et al. (1995) in which an attempt is made to to
explain in a coherent fashion the pattern of spectro-photometric properties of
elliptical galaxies, i.e. the CMR, the origin of the UV excess and its
dependence on the Mg2 index and total luminosity, etc. The models in question
differ for the kind of description adopted to follow the chemical history of
the galaxy: the one-zone closed-box scheme in Bressan et al. (1994) and the
one-zone infall scheme in Tantalo et al. (1995), this latter aimed at
simulating in a very simple fashion the formation of a galaxy  by collapse of 
primordial gas (cf. Chiosi 1981 for more details).

\littleskip
\noindent
(1) 
Elliptical galaxies are supposed to be made of two components, i.e.
luminous material of mass \ML\ embedded in a halo of dark matter of mass \MD,
whose presence affects only the gravitational potential of the galaxy and the
binding energy of the gas. The key assumption of the closed-box models is that
at the start of the star formation process (at time t=0) all the luminous mass 
is present in form of gas. In contrast, in the infall models the mass  of the
luminous component (supposedly  in form of gas) is let increase with time
according to 

\begin{equation}
  {d\ML(t) \over dt} = {\rm \dot M_{0}} ~exp(-t/\tau)  
\label{rate_infall}
\end{equation}
where $\tau$ is the accretion time scale. The constant ${\rm \dot M_0}$ is
obtained from imposing that at the galaxy age $T_{G}$  a certain value of
$\ML{\rm (T_{G})}$ is reached. Therefore, the time dependence of $\ML(t)$ is 

\begin{equation}
 {\rm \ML(t) =  { \ML(T_{G}) \over {[1-exp (-T_{G}/\tau)] } } 
                      [1 - exp(-t/\tau)]  }  
\label{asymptotic_mass}
\end{equation}
\noindent
As far as the dark component  $\MD$ is concerned, we simply assume it to be
constant in time and equal to a certain fraction of the asymptotic value of the
luminous mass (${\rm \ML(T_{G})/\MD}=\beta$), and to obey the spatial
distribution with respect to \ML\ given by  the model of Bertin et al. (1992)
and Saglia et al. (1992), according to whom  good fits of  elliptical galaxies
are possible  for  ratios  $\ML/\MD$ and $\RL/\RD$ equal to $\beta=0.2$. With
this assumption  little effects are to be expected from the presence of dark
matter. 
\littleskip

\noindent
(2) 
The stellar birth rate. i.e. the number of stars of mass M born in the time
interval dt and mass interval dM is 
\begin{equation}
    d{\cal N} =   \Psi(t, Z) \phi(M) dt dM 
\label{birth}
\end{equation} 
where per $\Psi(t)$ is the rate of star formation as a function of time and
chemical enrichment, while $\phi(M)$ is the initial mass function (IMF). 
The rate of star formation $\Psi(t)$ is taken  to be proportional to
the available gas mass $M_g(t)$, i.e. 
\begin{equation}
 \Psi(t)= \nu Mg(t)^k ~~~ \Msun~ {\rm yr^{-1} } 
\label{rate}
\end{equation}
with $k=1$. The constant of proportionality $\nu$ represents the inverse star
formation time scale in suitable units. 
The IMF is the Salpeter law
\begin{equation}
\phi(M)=   M^{-x} 
\label{phi_m}
\end{equation}
with $x=2.35$. The IMF is normalized by choosing the
 parameter $\zeta$ 

\begin{equation}
\zeta = {  { \int_{M^*}^{M_U}\phi(M){\times}M{\times}dM  } \over
 { \int_{M_L}^{M_U}\phi(M){\times}M{\times}dM  } }   
\label{phi_m_nor}
\end{equation}
fixing the fraction of  total mass  in the IMF above $ M^{*}$ and deriving 
the lower limit of integration $M_L$. The upper
limit of integration is $M_{U}=120 M_{\odot}$, the maximum mass in our data
base of stellar models, while the mass limit  $M^*$ is the  minimum mass
contributing to the chemical enrichment of the interstellar medium over a time
scale comparable to the total lifetime of a galaxy. This mass is approximately
equal to $1 M_{\odot}$. 
\littleskip

\begin{table}
\caption{Key parameters of the infall models and correspondence between \ML\
and  present-day total mass of stars \MS. Both are in units of  $10^{12}
\Msun$. The time scale of mass accretion,  $\tau$,  and the age at which
galactic winds occur,  $t_{gw}$, are expressed in  Gyr. } 
\littleskip
\begin{center}
\begin{tabular*}{80mm}{ccccccc} 
\hline 
\hline
\ML   &$\tau$ &$\nu$  & $t_{gw}$&   $Z$  & $<Z>$    & \MS      \\
\hline
\multispan{7}{ $\zeta=0.50$\hfil }\\
\hline
3.0  & 0.10  & 12.0   & 0.26   & 0.0710   & 0.0360  & 2.088    \\
1.0  & 0.10  & 7.2    & 0.31   & 0.0629   & 0.0307  & 0.648    \\
0.5  & 0.10  & 5.2    & 0.35   & 0.0543   & 0.0265  & 0.295    \\
0.1  & 0.10  & 3.0    & 0.34   & 0.0328   & 0.0166  & 0.041    \\
0.05 & 0.10  & 2.5    & 0.29   & 0.0235   & 0.0122  & 0.016    \\
0.01 & 0.10  & 1.0    & 0.43   & 0.0158   & 0.0080  & 0.002    \\
\hline 
\hline
\end{tabular*}
\end{center}
\label{tab_mass}
\end{table}

\noindent
(3) 
The detailed treatment of the chemical enrichment, i.e. without instantaneous
recycling (cf. Tinsley 1980),  is included so that the temporal evolution of
chemical abundances is properly taken into account (see Bressan et al. 1994 for
more details on the equations governing the chemical evolution of the models). 
\littleskip

\noindent
(4) 
The models include the presence of galactic winds triggered by the energy
deposit into the interstellar gas  due to the explosion of type I and II
supernovae, and also stellar winds from massive stars. When the energy storage
overwhelms the gravitational binding energy of gas this is supposed to be
expelled from the galaxy thus halting further star formation. The prescription
is exactly the same as in Bressan et al. (1994). 
\littleskip

\noindent
(5) 
The models are identified  by the mass of the luminous component \ML\ in units
of $10^{12}$ \Msun (thereinafter referred to as \M12). We remind the reader
that this mass never coincides with the present day mass in stars. In the
closed-box models \ML\ is the initial value of the luminous mass, part of which
is converted into stars and part is expelled by galactic winds. Furthermore,
the mass in living stars, \MS, most contributing to the light emitted by a
galaxy is subjected to decrease with time because of the continuous death of
stars leaving collapsed remnants and gas. In the case of infall models \ML\ is
the asymptotic value of the luminous mass which is never reached because
galactic wind is supposed to stop  star formation and gas accretion  at the
same time (cf. Tantalo 1994, and Tantalo et al. 1995). The same arguments for
the mass in stars given for the closed box models  apply also to the infall
models. In Table~\ref{tab_mass} we give the correspondence between \ML\ and 
\MS\ at the present age (15 Gyr). 

\noindent
(6) 
Finally, the isochrones at the base of the population synthesis calculations
are from the Padua library (Bertelli et al.  1994). Ample ranges of chemical
composition are considered, whose  parameters Y and Z  obey the enrichment law
\dydz=2.5 (cf. Pagel 1989, Pagel et al. 1992). The main assumptions to be
recalled here in view of the discussion below are (i) the 
use of the solar  pattern of abundances
for the so-called $\alpha$-elements with respect to Fe, [$\alpha$/Fe]=0; 
(ii) the use of $\eta=0.45$ in the mass loss rate along the RGB and AGB 
phases (cf. Bressan et al. 1994, Tantalo et al. 1995 for all details).
\littleskip
      
\begin{table*}
\caption{Integrated broad-band colours of infall models with \dydz=2.5 }
\smallskip
\begin{center}
\begin{tabular*}{135mm}{ccccccccc} 
\hline 
\hline
\ML   &  Age &  \Mbol  &    \Mv  &   \UB &   \BV &   \VR & \VK  & \UVex\\
\hline
     &     &        &        &      &      &      &      &      \\
 3.00& 17.0& -24.514& -23.662& 0.604& 0.999& 0.742& 3.330& 2.069\\
 3.00& 15.0& -24.597& -23.744& 0.597& 1.001& 0.743& 3.339& 2.217\\
 3.00& 12.0& -24.766& -23.916& 0.553& 0.981& 0.732& 3.347& 2.288\\
 3.00& 10.0& -24.972& -24.110& 0.525& 0.968& 0.727& 3.375& 2.155\\
 3.00&  8.0& -25.146& -24.315& 0.483& 0.937& 0.711& 3.339& 2.632\\
 3.00&  5.0& -25.560& -24.741& 0.444& 0.911& 0.691& 3.348& 6.093\\
     &     &        &        &      &      &      &      &      \\
 1.00& 17.0& -23.282& -22.470& 0.594& 1.000& 0.735& 3.273& 2.715\\
 1.00& 15.0& -23.369& -22.556& 0.586& 1.002& 0.736& 3.280& 2.938\\
 1.00& 12.0& -23.542& -22.733& 0.547& 0.986& 0.726& 3.289& 3.160\\
 1.00& 10.0& -23.738& -22.919& 0.522& 0.973& 0.720& 3.316& 3.070\\
 1.00&  8.0& -23.918& -23.122& 0.479& 0.946& 0.705& 3.287& 3.536\\
 1.00&  5.0& -24.332& -23.551& 0.420& 0.908& 0.682& 3.280& 6.068\\
     &     &        &        &      &      &      &      &      \\
 0.50& 17.0& -22.448& -21.668& 0.573& 0.994& 0.729& 3.222& 3.396\\
 0.50& 15.0& -22.537& -21.758& 0.566& 0.997& 0.729& 3.225& 3.792\\
 0.50& 12.0& -22.714& -21.940& 0.531& 0.983& 0.720& 3.234& 4.595\\
 0.50& 10.0& -22.901& -22.118& 0.507& 0.971& 0.713& 3.258& 4.712\\
 0.50&  8.0& -23.084& -22.320& 0.465& 0.946& 0.699& 3.233& 5.096\\
 0.50&  5.0& -23.498& -22.751& 0.396& 0.902& 0.674& 3.217& 6.019\\
     &     &        &        &      &      &      &      &      \\
 0.10& 17.0& -20.362& -19.652& 0.472& 0.947& 0.706& 3.080& 3.359\\
 0.10& 15.0& -20.454& -19.752& 0.470& 0.955& 0.706& 3.071& 3.887\\
 0.10& 12.0& -20.639& -19.945& 0.445& 0.946& 0.698& 3.073& 5.402\\
 0.10& 10.0& -20.802& -20.110& 0.420& 0.931& 0.689& 3.078& 5.937\\
 0.10&  8.0& -20.994& -20.314& 0.384& 0.909& 0.677& 3.064& 6.061\\
 0.10&  5.0& -21.398& -20.730& 0.321& 0.862& 0.652& 3.051& 5.804\\
     &     &        &        &      &      &      &      &      \\
 0.05& 17.0& -19.337& -18.675& 0.416& 0.915& 0.691& 2.977& 3.243\\
 0.05& 15.0& -19.433& -18.780& 0.418& 0.928& 0.692& 2.969& 3.818\\
 0.05& 12.0& -19.620& -18.974& 0.397& 0.922& 0.686& 2.970& 5.373\\
 0.05& 10.0& -19.778& -19.137& 0.373& 0.908& 0.677& 2.970& 5.915\\
 0.05&  8.0& -19.979& -19.344& 0.342& 0.887& 0.666& 2.963& 5.993\\
 0.05&  5.0& -20.359& -19.741& 0.279& 0.836& 0.639& 2.941& 5.628\\
     &     &        &        &      &      &      &      &      \\
 0.01& 17.0& -17.282& -16.704& 0.329& 0.860& 0.664& 2.789& 3.037\\
 0.01& 15.0& -17.393& -16.819& 0.335& 0.880& 0.667& 2.793& 3.751\\
 0.01& 12.0& -17.580& -17.014& 0.317& 0.879& 0.664& 2.794& 5.381\\
 0.01& 10.0& -17.740& -17.176& 0.298& 0.867& 0.656& 2.798& 5.805\\
 0.01&  8.0& -17.957& -17.393& 0.275& 0.850& 0.647& 2.803& 5.792\\
 0.01&  5.0& -18.307& -17.768& 0.214& 0.792& 0.616& 2.757& 5.225\\
\hline 
\hline
\end{tabular*}
\end{center}
\label{tab_col}
\end{table*}

\subsection{Main results for infall models} 
For the reasons amply discussed by Bressan et al. (1994) and Tantalo et al.
(1995) infall models  match the region 2000 - 3500 \AA\ of the spectrum of an
elliptical galaxy much better than the closed-box ones. Therefore in the
analysis below we adopt the infall models of Tantalo et al. (1995). They 
successfully reproduce the CMR for Virgo and Coma galaxies by Bower et al.
(1992a,b) and the UV properties of elliptical galaxies 
for the following  parameters: enrichment law \dydz=2.5,
infall time scale $\tau=0.1$ Gyr, $k=1$,  $\zeta=0.50$, and $\nu$ decreasing
from 12 for the 3 \M12\ galaxy down to 2.5 for the 0.05 \M12\ object. 

The main  properties of these models are  presented in  Table~\ref{tab_mass}
which contains the specific star formation efficiency $\nu$, the age at the
onset of the  galactic wind, the maximum metallicity Z, the mean metallicity
${\rm < Z >}$, and the mass \MS\ in stars at the present epoch. 

\begin{figure}
\psfig{file=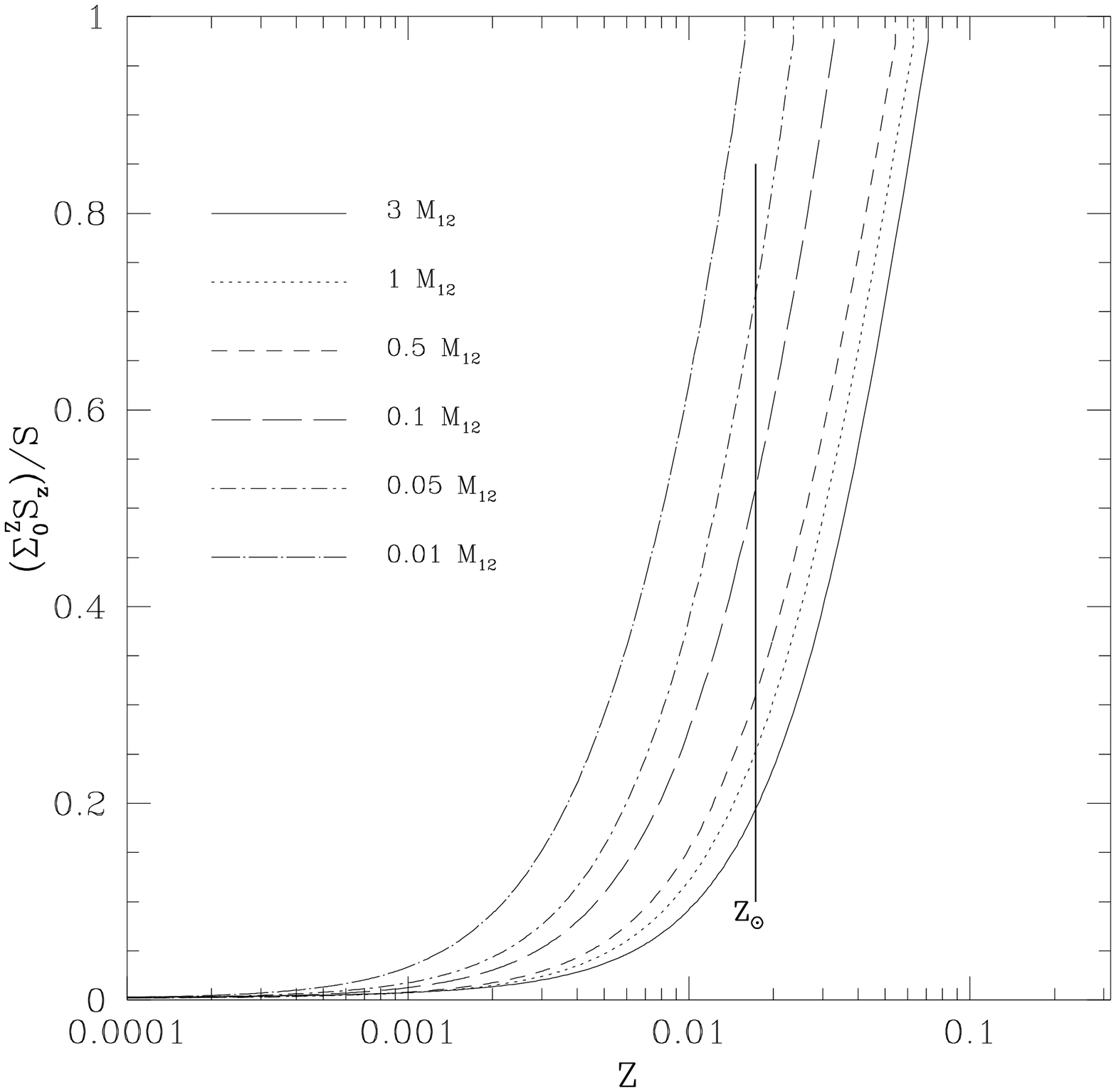,height=8.5truecm,width=8.5truecm}
\caption{Chemical structure of the galactic models with infall and different
\ML\ as indicated. The curves are the cumulative fractionary mass of living
stars per metallicity bin normalized to the present-day value of mass in stars
($\rm M_S$). The parameters of the models are those of Table 2. See the text
for more details} 
\label{chem_mod}
\end{figure}

The chemical structure of the models is represented by the 
normalized, cumulative
distribution of the mass in stars per metallicity bin at at the present epoch
(say 15 Gyr)  displayed in Fig.~\ref{chem_mod}. It is worth noticing the very
small percentage of stars of low metallicity as compared to the case of the
closed-box models of Bressan et al. (1994).

A summary of the temporal evolution of absolute bolometric and visual
magnitudes, ${\rm M_{bol} }$ and \Mv\ respectively, the broad-band colours \UB,
\BV, \VR, \VK, and   colour excess \UVex\ is given in Table~\ref{tab_col} for
selected values of the age. The CMR of these models is shown in Fig.~\ref{cmr}
for various values of the age as indicated. 

Like in Bressan et al. (1994) UV excess is generated by a suitable admixture of
three components (see also Greggio \& Renzini 1990). 
\littleskip
\noindent
(1) 
The classical post asymptotic giant branch (P-AGB) stars (see Bruzual 1992,
Bruzual \& Charlot 1993, Charlot \& Bruzual 1991) P-AGB stars are always
present in the stellar mix of a galaxy. The major problem with these stars is
their high effective temperature  and the  relation between their mass and that
of the progenitor. The initial-final mass relationship is not firmly
established. The most popular empirical determination is by Weidemann (1987).
Matching this relation depends on the efficiency of mass loss during the AGB
phase and other details of model structure (cf. Chiosi et al. 1992). 
\littleskip

\noindent
(2)
The hot horizontal branch (H-HB) and AGB-manqu\'e stars of very high
metallicity (say $Z \simeq 0.07$). which are expected to be present albeit in
small percentages in the stellar content of  bulges and elliptical galaxies in
general. Indeed, these stars have effective temperatures in the right interval
and generate ISEDs whose intensity drops short-ward of  about 1000 \AA\ by the
amount indicated by the observational data. (Ferguson et al. 1991, Ferguson \&
Davidsen 1993). The formation of these stars can occur either with low values
of the enrichment ratio ($\Delta Y/\Delta Z \simeq 1$) and strong dependences
of the mass-loss rates on the metallicity (Greggio \& Renzini 1990) or even
with canonical mass-loss rates and suitable enrichment laws ($\Delta Y/\Delta Z
\simeq 2.5$) as in  Horch et al. (1992) and Fagotto et al. (1994a,b,c). 
\littleskip

\noindent
(3) 
Finally, the very blue HB stars of extremely low metallicity (Lee 1994).  These
stars have effective temperatures  hotter than  about 15,000 K but much cooler
than those of the P-AGB stars. Therefore, depending on their actual effective
temperature, they can generate ISEDs in agreement with the observational data.
However, most likely they are not  the dominant source of the UV flux because
the analysis by Bressan et al. (1994) clarifies that  in the wavelength
interval $2000 < \lambda < 3000$ \AA\ the ISEDs of the  bulge of M31 and of
elliptical galaxies like NGC~4649 are fully consistent with the notion that
virtually  no stars with metallicity lower than $Z=0.008$ ought to exist in the
mix of these stellar populations. 

\begin{figure}
\psfig{file=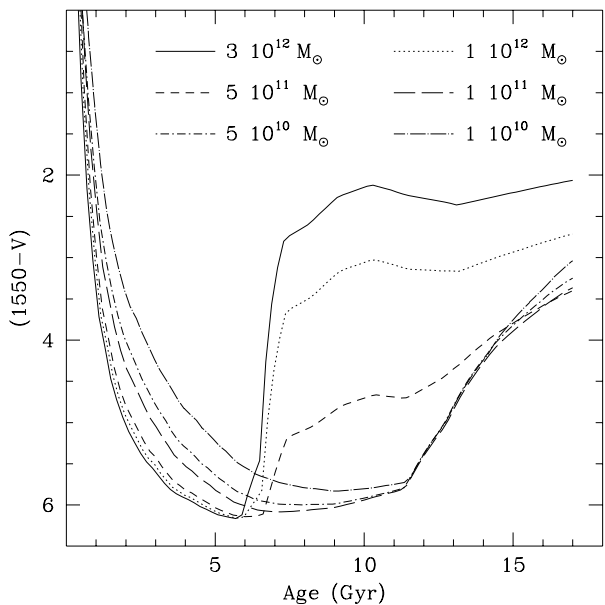,height=8.5truecm,width=8.5truecm}
\caption{ The time variation of the \UVex\ colour of the galactic 
models with infall and different
\ML\ as indicated. 
The parameters of the models are those of Table 2. See the text
for more details} 
\label{uv_excess}
\end{figure}

Our model galaxies with \dydz=2.5 emit UV radiation by old H-HB stars only for
ages older than about 5.6 Gyr and in presence of even a tiny fraction of 
stars in suitable metallicity bins (say $Z \geq 0.07$). The
precise value of the age depends also on the final-initial mass relationship
for the emitting stars (cf. Bressan et al. 1994, Tantalo et al. 1995 for
details). Fig.~\ref{uv_excess} show the \UVex\ colour of our galactic models 
as a function of the age. It is worth pointing out the different behaviour 
of the \UVex\ colour passing from the 3 \M12\ galaxy, in which H-HB,
AGB-manqu\'e, and P-AGB stars concur to generate the UV flux, to the 0.01 
\M12 galaxy, in which only P-AGB stars produce the  UV radiation. This trend is
caused by the decreasing mean and maximum metallicity at decreasing galaxy 
mass (cf. Bressan et al. 1994 for details).

Concluding this section, it is worth  pointing out that in our models the
efficiency of star formation per unit mass of gas increases at increasing
galactic mass ($\nu$ goes from 2.5 to 12.0 as the mass increases from 0.05 to 3
\M12). This is the reason for the nearly constant value of age  $t_{gw}$ at
which galactic winds occur halting star formation. Similar trend for the
parameter $\nu$ has been invoked by Matteucci (1994) to explain the abundance
ratios [Mg/Fe] observed in elliptical galaxies (see below).

\subsection{General remarks on the models}

We like to call the attention on three somewhat critical 
aspects of the models that could be used to invalidate the present
analysis. The points in question are  (i) the adoption of the one-zone
description which does not allow us to take into account spatial gradients in
colours and metallicity, (ii) the solar partition of abundances used in the
stellar models,  and finally (iii) our  treatment of galactic winds. 

(i) 
It is clear that the  \VK\ colours and magnitudes of the galaxies in the Bower
et al. (1992a,b) sample  refers to the whole galaxy, whereas the UV excess and
its companion colour \UVex\ of Burstein et al. (1988) in most cases refer to
the central region of a galaxy. This is also true for the various indices
measured by Gonzales (1993) to be  discussed below. The existence of gradients
in colours and metallicities across elliptical galaxies is a well established
observational fact (Carollo et al. 1993; Carollo \& Danziger 1994, Davies et
al. 1993, Schombert et al. 1993). This means that insisting on the one-zone
description as in the present models, could be source of difficulty when
comparing model results with observational data. It seems reasonable to argue
that higher metallicities can be present in the central regions thus
facilitating  the formation of the right type of stellar sources of the UV
radiation (H-HB and AGB manqu\'e stars), whereas lower metallicities across the
remaining parts of the galaxies would make the other integrated colours such as
(V-K) bluer than expected from straight use of the one-zone model for the whole
galaxy.

(ii) 
Recent observations indicate that the pattern of abundances in elliptical
galaxies is skewed towards an overabundance of $\alpha$-elements with respect
to Fe (cf. Carollo et al. 1993; Carollo \& Danziger 1994). As already recalled,
 the library of stellar models in use is based on the standard (solar) pattern
of abundances.  Work is in progress to generate libraries of stellar models 
and isochrones with $[\alpha/{\rm Fe}] > 0$.   
Preliminary calculations for the solar metallicity (Z=0.02) show the effect 
to be small in 
the theoretical CMD. In addition, calculations of the Mg2 index from 
synthetic spectra  for different metallicities and partitions 
of $\alpha$-elements (Barbuy 1994) 
show that  passing from $[\alpha/{\rm Fe}=0.0$ to 
$[\alpha/{\rm Fe}=0.3$ the Mg2 index of an old SSP (15 Gyr)
increases from 0.24 to 0.30. 
\littleskip

(iii)
Concerning galactic winds, we would like to  comment   on the recent claim by
Gibson (1994) that Bressan et al. (1994) underestimate the effect of supernova
explosions and on the contrary overestimate the effect of stellar winds from
massive stars. In brief,  Bressan et al. (1994) assuming  the CMR to be 
a mass-metallicity sequence
looked at the metallicity that would
generate the right colours.  They found that considering supernovae as the only
source of energy, by the time the galactic winds occur,  the gas fraction has
become too low and the metallicity too high in turn, that the CMR  is destroyed
(it runs flat). To cope with this difficulty they included another source of
energy, i.e. the stellar winds from massive stars. In analogy with the
behaviour of a SN remnant's thermal energy, which decreases with time as
$\epsilon_{SN} \propto (t/t_c)^{-0.62}$  for times greater than $t_c$,  Bressan
et al. (1994) assumed that the energy  of stellar winds obeys the same law but
with a different $t_c$, i.e. $\epsilon \propto (t/t_{cw})^{-0.62}$ for times
greater than $t_{cw}$. The time $t_{cw}$ was considered as a parameter. They
found that $t_{cw}=1.5\times 10^7$ yr is a good choice.  This time  roughly
corresponds to the evolutionary lifetime of  a 10 \Msun\ star. In other words it
is the time scale over which a group of newly formed  O-type stars would evolve
away from the SSP. They found that $t_{cw}$ shorter than this would not allow
sufficient powering of the interstellar medium, so that galactic winds  would
occur much later than required by the CMR. Bressan et al. (1994) provided also
an explanation for the different results found by other authors (Arimoto \&
Yoshii 1987, Matteucci \& Tornambe' 1987, Angeletti \& Giannone 1990, and
Padovani \& Matteucci 1993).  Admittedly, this additional source of energy was
invoked to keep the standard interpretation of the CMR.

\section{Narrow-band indices for SSPs and galaxies }
In this paper we make use of the recent library of empirical calibrations of
twenty one indices of absorption feature strengths calculated by  Worthey
(1992) and Worthey et al. (1994). The calibrations are presented in form of
analytical fits that give the index strength as a function of stellar
atmospheric parameters, i.e. $\Theta_e =5040/\Teff$, $\log g$ (gravity),  and
\FeH\ both for warm (up to 13,260 K) and cool stars (down to 3570 K). The
definition of the indices are given in Worthey (1992) and Worthey et al. (1994)
to whom we refer. Suffice the recall here that the index \MgFe\ is the
geometric mean of the two indices $\rm Mgb$ and \MFe, \MgFe\ = $({\rm
Mgb}~\MFe)^{0.5}$ where in turn the index \MFe\ is the mean of the indices $\rm
Fe_{5270}$ and $\rm Fe_{5335}$, \MFe=($\rm Fe_{5270}$ +  $\rm Fe_{5335}$)/2. 

According to their definition, all  indices are  constructed by means of a
central band-pass and two pseudo-continuum band-passes on either side of the
central band (cf. Worthey et al. 1994 for details). The continuum flux is
interpolated between the mid points of the pseudo-continuum band-passes. 

The integrated indices for SSPs and model galaxies are calculated using the
following method. 

For every  combination of \Teff, g and \FeH, first  we derive  the flux in the
continuum band-pass  from the library of stellar spectra used by Bressan et al.
(1994) implemented by the revision of Tantalo et al. (1995), and then we
calculate the  flux in the central pass-bands with  the analytical fits of
Worthey et al. (1994) for the same values \Teff, g and \FeH. Let us call
$F_{I\lambda}$ and $F_{C\lambda}$ the flux in central pass-band and in the
continuum, respectively. 

The  integrated  $F_{I\lambda}$ and $F_{C\lambda}$ (thereinafter indicated as
$F_{\kappa}$)  for the  stellar content of a galaxy of age $T$ are given by 

\begin{equation}
F_{\kappa}(T) = \int_0^T \int_{M_L}^{M_U} \Psi(t,Z)\phi(M)\ F_{\kappa}
(M,\tau',Z) \ dt \ dM 
\label{ind_1}
\end{equation}
where $F_{\kappa}(M,\tau',Z)$ is one of the two fluxes for a star of mass $M$,
metallicity $Z(t)$, and age $\tau'=T-t$. Separating the contribution from
single SSPs, the above integral becomes 

\begin{equation}
F_{\kappa}(T) = \int_0^T \Psi(t,Z)\ {\it F_{\kappa, ssp} }(\tau',Z)\ dt 
\label{ind_2}
\end{equation}
where 

\begin{equation}
{\it F_{\kappa, ssp} }(\tau',Z) = 
\int_{M_L}^{M_U} \phi(M)\ F_{\kappa}(M,\tau',Z)\ dM 
\label{ind_3}
\end{equation}
are  defined as the integrated fluxes, either $F_{I\lambda}$ or $F_{C\lambda}$
for a single SSP,  i.e. of a coeval, chemically homogenous assembly of stars
with age $\tau'$ and metallicity Z. 

Knowing the integrated fluxes either for a SSP or a galaxy, the definition of
each index is applied to get back the integrated index.

The  rate of star formation $\Psi(t,Z)$ and initial mass function $\phi(M)$ in
equation (\ref{ind_2}) are the same as in  equations (\ref{rate}),
(\ref{phi_m}), and (\ref{phi_m_nor}).


\setcounter{table}{4}
  
\begin{table*}
\caption{Theoretical indices for model galaxies with different 
\ML\ and age (in Gyr).  }
\littleskip
\begin{center}
\begin{tabular*}{140mm}{ccccc c ccccc } 
\hline 
\hline
\ML &  Age  &  $\log\Hbeta$ &   $\log\MgFe$  & $\log$Mg2&  &
      \ML &  Age  &  $\log\Hbeta$ &   $\log\MgFe$  & $\log$Mg2  \\
\hline
3.00&   17.0&   0.157&    0.552& 0.277&& 3.00& 10.0&  0.204&    0.533& 0.261 \\ 
1.00&   17.0&   0.167&    0.536& 0.265&& 1.00& 10.0&  0.209&    0.517& 0.250 \\
0.50&   17.0&   0.177&    0.520& 0.254&& 0.50& 10.0&  0.215&    0.500& 0.239 \\
0.10&   17.0&   0.211&    0.468& 0.221&& 0.10& 10.0&  0.241&    0.446& 0.205 \\
0.05&   17.0&   0.234&    0.433& 0.201&& 0.05& 10.0&  0.256&    0.412& 0.186 \\
0.01&   17.0&   0.278&    0.368& 0.170&& 0.01& 10.0&  0.284&    0.349& 0.157 \\
    &       &        &         &      &&     &     &       &        &       \\ 
3.00&   15.0&   0.155&    0.551& 0.277&& 3.00&  8.0&  0.245&    0.518& 0.249 \\
1.00&   15.0&   0.163&    0.535& 0.265&& 1.00&  8.0&  0.245&    0.502& 0.238 \\ 
0.50&   15.0&   0.172&    0.518& 0.253&& 0.50&  8.0&  0.248&    0.485& 0.228 \\ 
0.10&   15.0&   0.201&    0.466& 0.219&& 0.10&  8.0&  0.273&    0.430& 0.195 \\ 
0.05&   15.0&   0.220&    0.432& 0.200&& 0.05&  8.0&  0.289&    0.396& 0.177 \\
0.01&   15.0&   0.257&    0.369& 0.169&& 0.01&  8.0&  0.315&    0.334& 0.150 \\
    &       &        &         &      &&     &     &       &         &     \\
3.00&   12.0&   0.179&    0.539& 0.267&& 3.00&  5.0&  0.298&    0.496& 0.232 \\ 
1.00&   12.0&   0.185&    0.524& 0.255&& 1.00&  5.0&  0.308&    0.476& 0.220 \\ 
0.50&   12.0&   0.191&    0.508& 0.244&& 0.50&  5.0&  0.317&    0.457& 0.209 \\
0.10&   12.0&   0.216&    0.457& 0.212&& 0.10&  5.0&  0.345&    0.400& 0.179 \\ 
0.05&   12.0&   0.232&    0.423& 0.193&& 0.05&  5.0&  0.362&    0.362& 0.161 \\
0.01&   12.0&   0.261&    0.360& 0.163&& 0.01&  5.0&  0.394&    0.291& 0.133 \\
    &       &        &         &      &&     &     &       &         &       \\ 
\hline
\hline 
\end{tabular*}
\end{center}
\label{tab_theo_in}
\end{table*}

\begin{figure}
\psfig{file=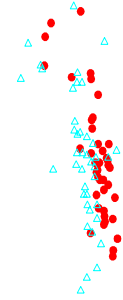,height=9.5truecm,width=8.5truecm}
\caption { Single Stellar Populations of different metallicity in the \Hbeta\ -
\MgFe\   plane. Along each the age varies from 1  to above 15 Gyr as indicated.
The full dots and open triangles are the Re/8-data and Re/2-data of  Gonzales
(1993). The error bars of the data are those of 
Fig.~\protect\ref{gonz_data} } 
\label{ssp}
\end{figure}

Tabulations of indices for SSPs with different chemical composition and age are
given in Table~4. In order to present results independent of the particular
choice made for the IMF normalization, the SSPs of Table~4 are calculated
assuming the Salpeter IMF with $x=2.35$, and fixed lower and upper limits of
integration,  $M_L=0.15 M_{\odot}$ and  $M_U=120 M_{\odot}$, respectively.
Care must be paid applying these
indices for SSPs to real assemblies of stars (either clusters or galaxies). 
 
The correlation between \Hbeta\ and \MgFe\ for SSPs with different
metallicities is shown in Fig.~\ref{ssp}.  The age goes from 1 Gyr to beyond 15
Gyr in steps of  $\Delta\log t=0.05$ as indicated. The inspection of the data
in Table~4 and comparison with similar results obtained by Gonzales (1993) show
that excellent agreement exists between the two studies. The dip and the
loop-like structure in the SSPs with the lowest and highest metallicity are
caused by the onset of the HB and appearance of the H-HB  stars, respectively.
Only for the sake of provisional comparison we plot in Fig.~\ref{ssp} also the
Re/2-data (open triangles) and Re/8-data (filled circles)  of the Gonzales
(1993) sample. It is soon evident that the locus drawn by the observations has
the same slope of the SSPs with metallicity comprised between Z=0.02 and
Z=0.05.

\section{Elliptical galaxies in the \Hbeta\ - \MgFe\ plane}

With the aid of the technique described in the previous sections, 
we calculate the
temporal evolution of the narrow band indices for model galaxies with different
\ML. Table~\ref{tab_theo_in} lists  the indices \Hbeta\, \MgFe\, and Mg2 of our
 galactic models at selected values of the age, namely 17, 15, 12, 10, 8 and 5
Gyr.

\begin{figure}
\psfig{file=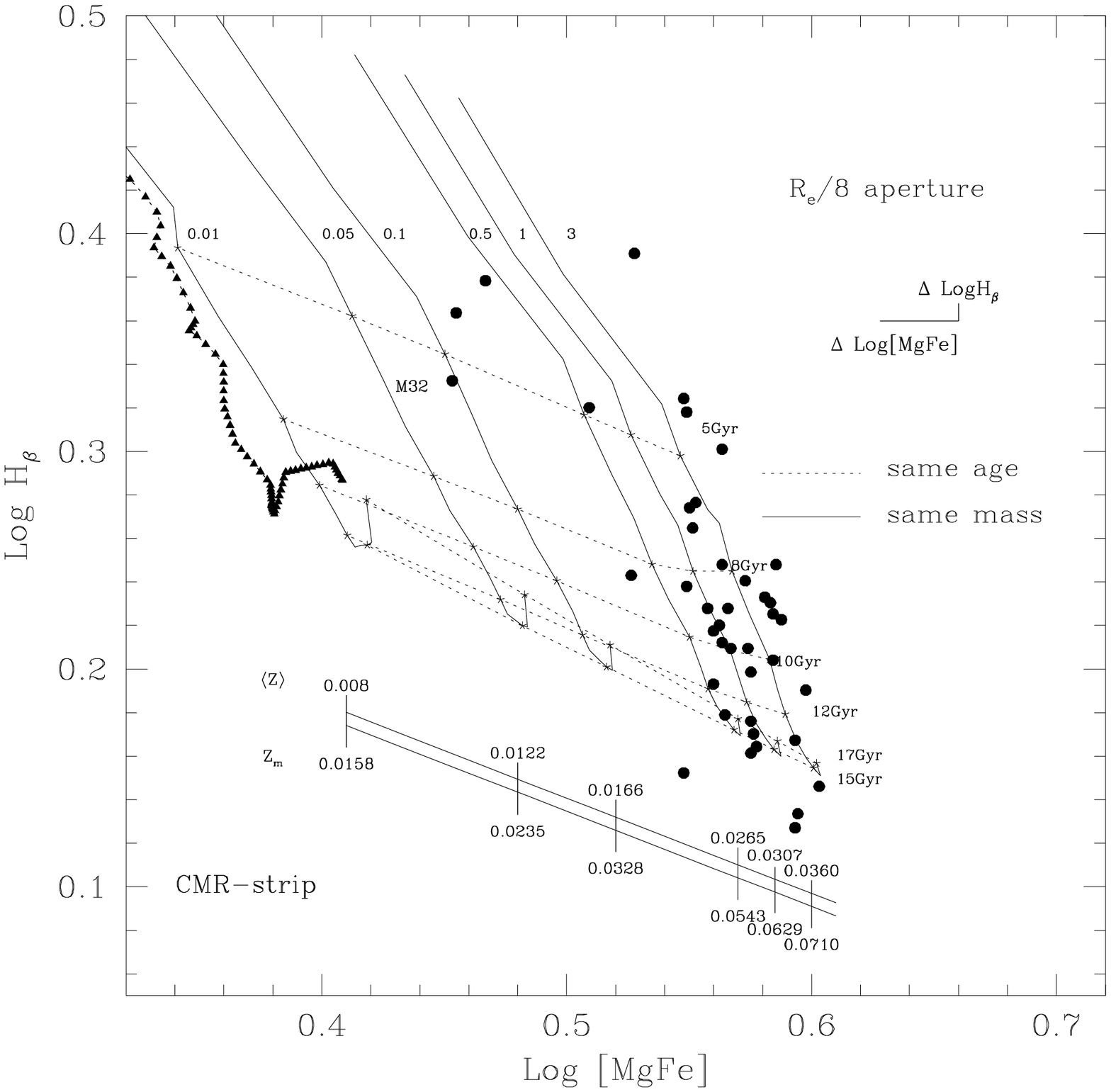,height=9.5truecm,width=8.5truecm}
\caption{Evolution of the infall  models in the \Hbeta\ - \MgFe\ plane (solid
lines). Each model  is labelled  by \ML. The dotted lines are isochrones of
different age (in Gyr) as indicated. The line called CMR-strip with the mean
and maximum metallicities of the galactic models annotated along it is the
analog  of the CMR. The solid triangles show the evolutionary path of the 3
\M12\ galaxy of Bressan et al. (1994) calculated with the closed-box formalism.
The full dots are the Re/8-data of Gonzales (1993). Finally, the vectors
$\Delta \log\Hbeta$ and $\Delta \log\MgFe$ display the mean shift in the data
going from the  R/8-data to the Re/2-data.  The error bars of the data are 
those of Fig.~\protect\ref{gonz_data} See the text for more details  } 
\label{gal_hb_mg}
\end{figure}

The evolutionary path of model galaxies in the  \Hbeta\ - \MgFe\  plane is
shown in Fig.~\ref{gal_hb_mg} together with the Re/8-data of the Gonzales
(1993) sample.   The galactic
models  are labelled by their \ML.  Along each sequence the age increases from
the top to the bottom, going from  1 to beyond 15 Gyr as indicated. 
Although the Re/2-data would indeed be closer to the
theoretical results for the one-zone models we are using, we prefer 
to adopt the Re/8-data referring to the central part of the galaxies.
However, we would like to clarify that similar results would be obtained
using the  Re/2-data. Indeed passing from Re/2 to Re/8 the net effect is 
that \Hbeta\ and \MgFe\ indices get slightly "redder" by 
the quantities $\Delta\log\Hbeta= 0.003$ and
$\Delta\log\MgFe=0.032$ as indicated by the vectors shown in
 Fig~\ref{gal_hb_mg}.

In any case, compared with the Re/8-data, our values for \MgFe\ are 
somewhat bluer than
observed.  This could reflect the solar partition of abundances adopted for the
stellar models in use. It could also mean  that the very central parts of the
galaxies are more metal-rich than predicted by the one-zone model. Indeed there
is much better  agreement between the Re/2-data and theoretical models.
Therefore, we consider the above discrepancy to be marginal and not
invalidating the analysis below. It will be taken into account whenever
necessary. On the basis of the above remarks, we consider it reasonable to
impose coincidence between our reddest models and the red edge of the
observational distribution, and apply to the models an  offset of
$\Delta\log\MgFe=0.05$. 
The required shift is comparable with the increase in the Mg2 index found by 
Barbuy (1994) for ${\rm [Mg/Fe]=0.3}$ and reported in section 3.2.
This shift will always be applied when comparing 
models with observations.
  The same is true for the Re/2-data. In such a case the
shift along the \MgFe\ axis amounts only to about $\Delta\log\MgFe=0.02$.

As expected the evolutionary paths of the galaxies run parallel to that of
SSPs. This is the consequence of interpreting the CMR as a mass-metallicity
sequence, along which both the mean and the maximum metallicity of a galaxy are
ultimately  determined by the onset of galactic winds. In the canonical view of
the CMR (slope), galactic winds and consequent interruption of the star forming
activity occur earlier  at decreasing galactic mass (lower mean and maximum
metallicities). Since the subsequent  evolution is passive at frozen chemical
composition, the behaviour of a SSP is mimicked. Therefore, in the  \Hbeta\
- \MgFe\ plane galactic models shift to the left at decreasing mass and mean
metallicity in turn, cf. the entries of Table~\ref{tab_mass} and the values
marked along the line of  Fig.~\ref{gal_hb_mg} labelled "CMR-strip" (see below)
which shows the mean slope of the old isochrones (10 to 17 Gyr). The mean
metallicity is annotated above, whereas the maximum metallicity is written
below the line in correspondence to the galactic models with different \ML. 

For the purposes of comparison, in Fig.~\ref{gal_hb_mg} we also show  the most
massive closed-box model of Bressan et al. (1994), namely the case with 3 \M12\
(triangles).  It is interesting to note the large effect passing from 
closed-box to infall models. At given \ML, the closed-box models  are much
bluer both in \Hbeta\ and \MgFe\ than the infall models. No galaxy is found in
the region occupied by the  closed-box  model with 3 \M12\ and lower. This is
another argument in favour of the  infall models. 
 
In order to quantify the effect of variations in metallicity and age on the
 position of a galaxy  in the  \Hbeta\ - \MgFe\ plane, we calculate the vectors
$\Delta\log\Hbeta$ and $\Delta\log\MgFe$  at constant age and constant
metallicity. To this aim, we consider models of given \ML\  to derive
$(\Delta\log\Hbeta)_t$ and $(\Delta\MgFe)_t$ and models of fixed age and
different \ML\ to get $(\Delta\log\Hbeta)_Z$ and $(\Delta\log\MgFe)_Z$, with
obvious meaning of the symbols. We find the following relations over wide
ranges of ages and metallicities 

\begin{equation}
(\Delta\log\Hbeta)_t = ({\partial\log\Hbeta \over \partial Z})_t \simeq -1.7
\end{equation}

\begin{equation}
(\Delta\log\Hbeta)_Z =({\partial\log\Hbeta \over \partial t})_Z \simeq -0.01 
                         ~~~Gyr^{-1}  
\end{equation}
and
\begin{equation}
(\Delta\log\MgFe)_t =({\partial\log\MgFe \over \partial Z})_t \simeq 3.5 
\end{equation}

\begin{equation}
(\Delta\log\MgFe)_Z =({\partial\log\MgFe \over \partial t})_Z \simeq 0.005
                           ~~~Gyr^{-1} 
\end{equation}

Looking at the data of Fig.~\ref{gal_hb_mg}, although some scatter is present
along the \MgFe\ axis, most of the galaxies fall  into the region 
between the sequences with  0.5 and  3 \M12. 
This implies that the metallicity distribution within most of the
 galaxies in the sample is quite similar with both the mean and maximum 
metallicity nearly identical  (see the metallicity distributions of the 0.5 
and \ML\ galaxies shown in  Fig.~\ref{chem_mod}).
However, there are a few
galaxies (M32 and NGC~7454, in particular)  that apparently fall along
sequences of smaller \ML\ and hence, in our scheme,   lower metallicities. 
 In contrast, there is a large scatter
along the \Hbeta\ axis that seems to hint  large variations in the age from
galaxy to galaxy. This indeed is the conclusion reached by Gonzales (1993). For
instance, the prototype galaxies M32 and NGC~4649 would turn out to be as young
as 3  Gyr and as old as 17 Gyr, respectively.

There is  a puzzling feature coming out from the close inspection of 
Fig.~\ref{gal_hb_mg}. In brief, as  the galactic
models have been constrained to match the slope and the \VK\ colours of the CMR
under the assumption that this latter a metallicity-mass sequence of old (say
15 Gyr), nearly coeval objects,  we  expect galaxies to be located in
the \Hbeta\ - \MgFe\ plane   along  a the line indicated as CMR-strip, which is
the analog of the CMR of Fig~\ref{cmr}. 
It is soon evident that the slope of the CMR-strip
does not agree with the slope traced by the observational data. 
One may argue
that this results from the lack of homogeneity between the samples of 
Gonzales (1993) and Bower et al. (1992a,b). Coevality holds for the cluster
galaxies of Bower et al. (1992a,b), whereas the spread in age applies to the
field galaxies of Schweizer \& Seitzer (1992) and Gonzales (1993).
In both cases, 
 as the CMRs of   cluster and field galaxies have similar slope,
 one would expect galaxies to scatter within the area comprised between the 
youngest and oldest isochrone pertinent to the sample.

In the scenario  of coeval, old galaxies, the four objects of  Bower
et al. (1992a,b) in common with Gonzales (1993) -- see the entries of
Table~\ref{tab_data} -- should
 fall in the red corner of the \Hbeta\ - \MgFe\ plane
with little dispersion. This is not true, because NGC~4478 has a significantly
bluer \Hbeta\ suggesting a younger age, much younger than the range permitted
by the tightness of the CMR. 
Conversely, in the scenario of galaxies with different ages (as perhaps 
induced by mergers), the associated chemical enrichment must be such that
galaxies acquire very similar metallicity distributions.

What we learn from this  preliminary analysis of the observed distribution
of galaxies in the  \Hbeta\ - \MgFe\ plane is: 

\begin{itemize}
\item{ In spite of their total luminosity and mass, most galaxies seem to
possess nearly identical chemical structures. } 

\item{ Ages seem to vary from galaxy to galaxy. }

\item{ Galaxies do not simply 
distribute along the locus expected for objects matching
the  mass-metallicity sequence of the  CMR. } 

\end{itemize}

The following questions can be  addressed:  Is the age spread real ? Does it
imply a truly different starting epoch of the star formation activity, or 
other effects  could change the evolutionary path of a galaxy in the   \Hbeta\
- \MgFe\ plane and mimic a spread in age ? Why the predictions from the CMR do
not find close correspondence in the \Hbeta\ - \MgFe\ plane ?

\begin{figure}
\psfig{file=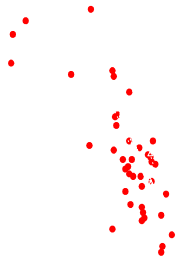,height=9.5truecm,width=8.5truecm}
\caption{Effects of a recent burst of star formation on the evolutionary path
in the \Hbeta\ - \MgFe\ plane.   The galaxy has mass of 3 \M12. The burst is
supposed to start at 12 Gyr with such an intensity that 1.0\% of the luminous
mass of the galaxy in form of gas is turned into stars. The full dots are the
Re/8-data of Gonzales (1993). The error bars of the data are those of 
Fig.~\protect\ref{gonz_data}   } 
\label{burst_mod}
\end{figure}

\begin{figure}
\psfig{file=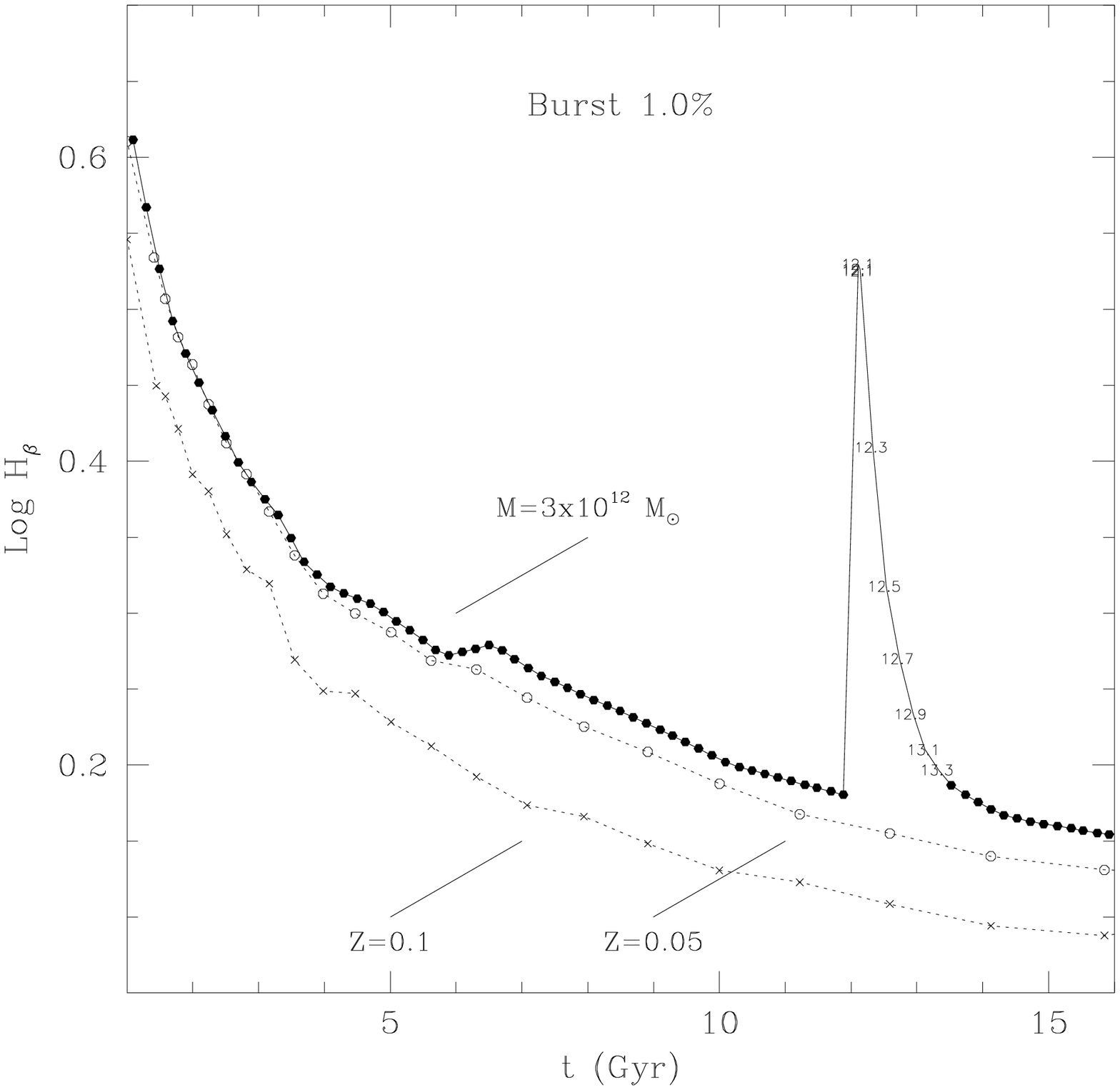,height=9.5truecm,width=8.5truecm}
\caption{The variation of the \Hbeta\ index as a function of time for the 
galaxy with a recent burst of star formation shown in
Fig.~\protect\ref{burst_mod} The thin lines show two SSPs with different 
metallicities as indicated  } 
\label{hb_burst}
\end{figure}

\section{Recent episodes of star formation ?}

As already recalled, a growing body of literature on fine structures and
kinematic anomalies suggest the early type galaxies were either formed or
restructured by mergers in relatively recent times (cf. Schweizer \& Seitzer
1992, Charlot \& Silk 1994, and references). In addition, studies of population
synthesis (Faber et al. 1992 and references) and  photometric data
(color-magnitude diagrams) of the stellar content in nearby galaxies (cf. the
case of M32 by Freedman 1989, 1992) seem to suggest that E+S0 galaxies contain
various admixtures of intermediate age stars. The merger hypothesis implies
star burst of various intensity (cf. the models of merging spirals by
Alfensleben \& Gerhard 1994), whose net result is the formation of an E/S0
galaxy. 
It is worth clarifying here that the term "merger"  may refer
 to  quite different 
views: either the classical merger of equal-mass spiral galaxies (Schweizer 
\& Seitzer 1992) or most of the stars forming at an early epoch and very 
modest star formation at later epochs (low red-shifts) as pointed out by 
Charlot \& Silk (1994), who apparently concluded that the data could not help 
to distinguish between these two scenarios.

How  a recent burst  of star formation, superposed to  much older populations,
would affect the evolutionary path of a galaxy in the \Hbeta\ - \MgFe\ plane is
easy to foresee by means of the following experiments. Having in mind the case
of M32 for which  the existence AGB stars as young as about $3\div 5$  Gyr
(Freedman 1989, 1992) has been suggested, we suppose that at the arbitrary age
of 12 Gyr a small burst of short duration occurs. The intensity of the burst is
such that only 1.0\% of the current luminous mass of the galaxy is turned into 
stars, whereas its duration is only  $1\times 10^8$ yr.  As expected and in
analogy to what is known from  similar experiments in the \UB\ and \BV\ colours
(cf. the models of Alfensleben \& Gerhard 1994), the occurrence of a  burst of 
star formation  induces a loop in the \Hbeta\ - \MgFe\ plane.  At the start of
the burst both \Hbeta \ and \MgFe\ get bluer, and  when the burst is over they 
recover the previous values. This is shown in Fig.~\ref{burst_mod} and
Fig.~\ref{hb_burst}. For this burst, the recovery time scale is about
1 Gyr. Therefore, in order to be detected a burst of this type should have
occurred as recently as 1 Gyr ago. Of course, the intensity and duration of the
burst may change from galaxy to galaxy, so that both the amplitude of the loop 
in the \Hbeta\ - \MgFe\ plane and the time scale at which the pre-burst values
are recovered may  be different from the ones we have been using. Stronger and
longer bursts induces larger amplitudes of the loop and somewhat longer
recovery time scales.

\begin{table*}
\caption{Mass in stars, \MS, radius of the luminous component \RL,
gravitational potential $\Omega$, and velocity dispersion $\Sigma$ of the
model galaxies. The units of each quantity are indicated. } 
\littleskip
\begin{center}
\begin{tabular*}{105mm}{ccccccc} 
\hline 
\hline
\ML\ & \MS     & \RL   & $\Omega$ & $\Sigma$   & $\log\Sigma$ & 
                                        $\log\Sigma_{ad}$ \\ 
\hline
$10^{12}\Msun$ & $10^{12}\Msun$ &  Kpc  &  ergs  & $Km sec^{-1}$  & & \\
\hline
3    & 2.08    & 39.1  & 6.18(60)  & 308  & 2.488 & 2.506 \\
     &         & 47.8  & 1.05(61)  & 334  & 2.523 &       \\
1    & 0.65    & 20.6  & 1.14(60)  & 237  & 2.374 & 2.395 \\
     &         & 26.1  & 2.13(60)  & 261  & 2.416 &       \\
0.5  & 0.29    & 13.3  & 3.64(59)  & 197  & 2.295 & 2.322 \\
     &         & 17.8  & 7.82(59)  & 223  & 2.348 &       \\
0.1  & 0.041   &  4.5  & 2.06(58)  & 127  & 2.104 & 2.148 \\
     &         &  7.3  & 7.58(58)  & 155  & 2.191 &       \\
0.05 & 0.016   &  2.7  & 5.32(57)  & 103  & 2.012 & 2.068 \\
     &         &  5.0  & 2.78(58)  & 133  & 2.123 &       \\
0.01 & 0.002   &  0.9  & 2.69(57)  &  64  & 1.80  & 1.88  \\
     &         &  2.1  & 2.61(56)  &  92  & 1.96  &       \\
\hline 
\hline
\end{tabular*}
\end{center}
\label{tab_sigma}
\end{table*}

On the basis of these experiments, one could argue that the large spread along
the \Hbeta\ direction is the result of a recent episode of star formation
(either triggered by a  merger or caused by an internal agent) superposed to an
older population whose properties are temporarily masked by the star forming
activity so that deciphering the true age of the system from the location in
the \Hbeta\ -  \MgFe\ plane is not possible. 

However, before accepting this conclusion, we feel it wise to examine in detail
the correlation of \Hbeta\ and  \MgFe\ with the colour \UVex\ and the velocity
dispersion  $\Sigma$.

\begin{figure}
\psfig{file=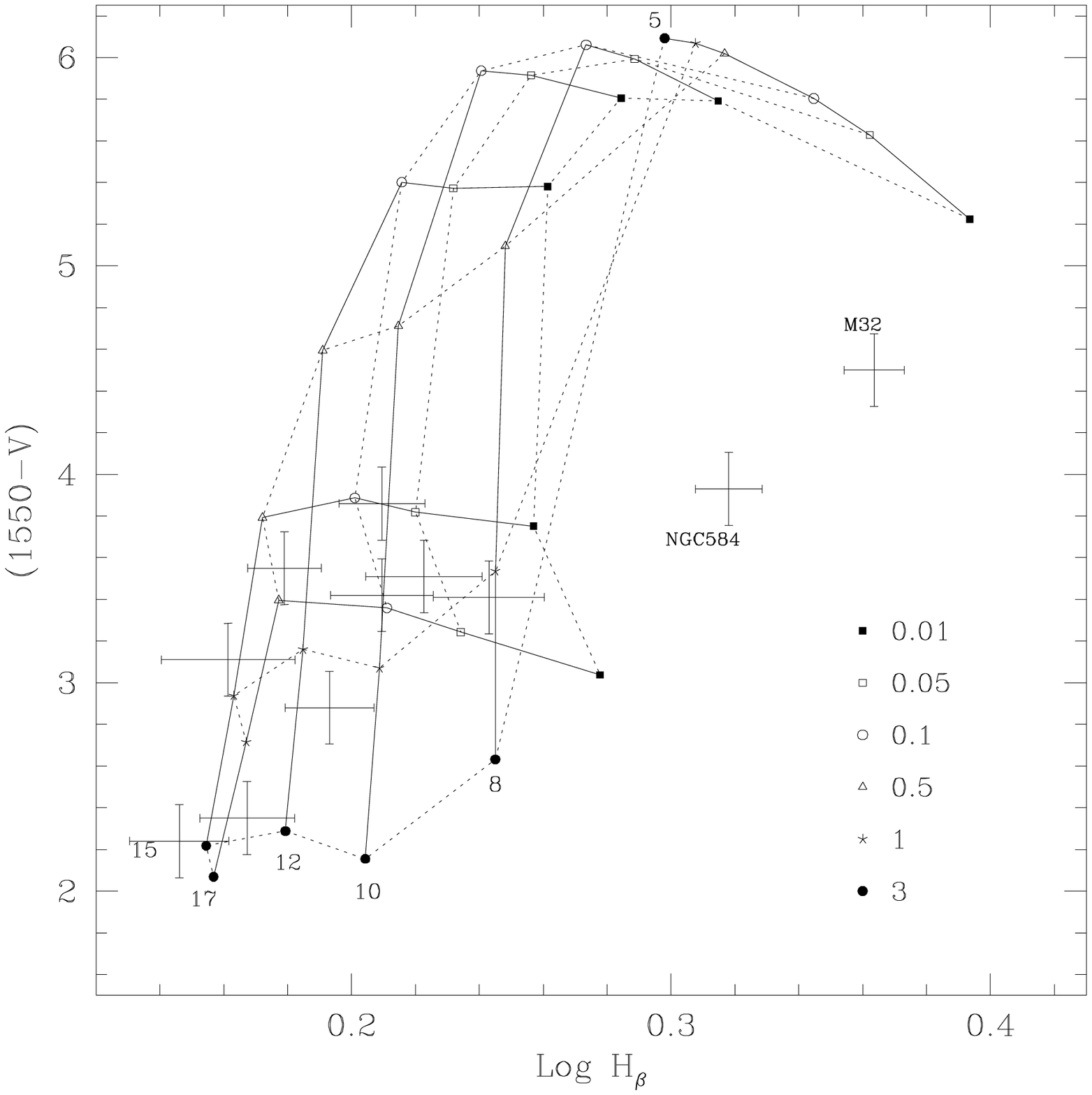,height=9.5truecm,width=8.5truecm}
\caption{ The relation between \UVex\ and \Hbeta. The indices \Hbeta\ are from
Gonzales (1993), whereas the colours \UVex\ are from Burstein et al. (1988).
The crosses show the uncertainty affecting the observational data. The
solid and dotted lines correspond to loci of constant age (in Gyr) and mass (in
units of \ML\  as indicated. The position of M32 and NGC~584 is brought into
evidence  } 
\label{uv_hb}
\end{figure}

\begin{figure}
\psfig{file=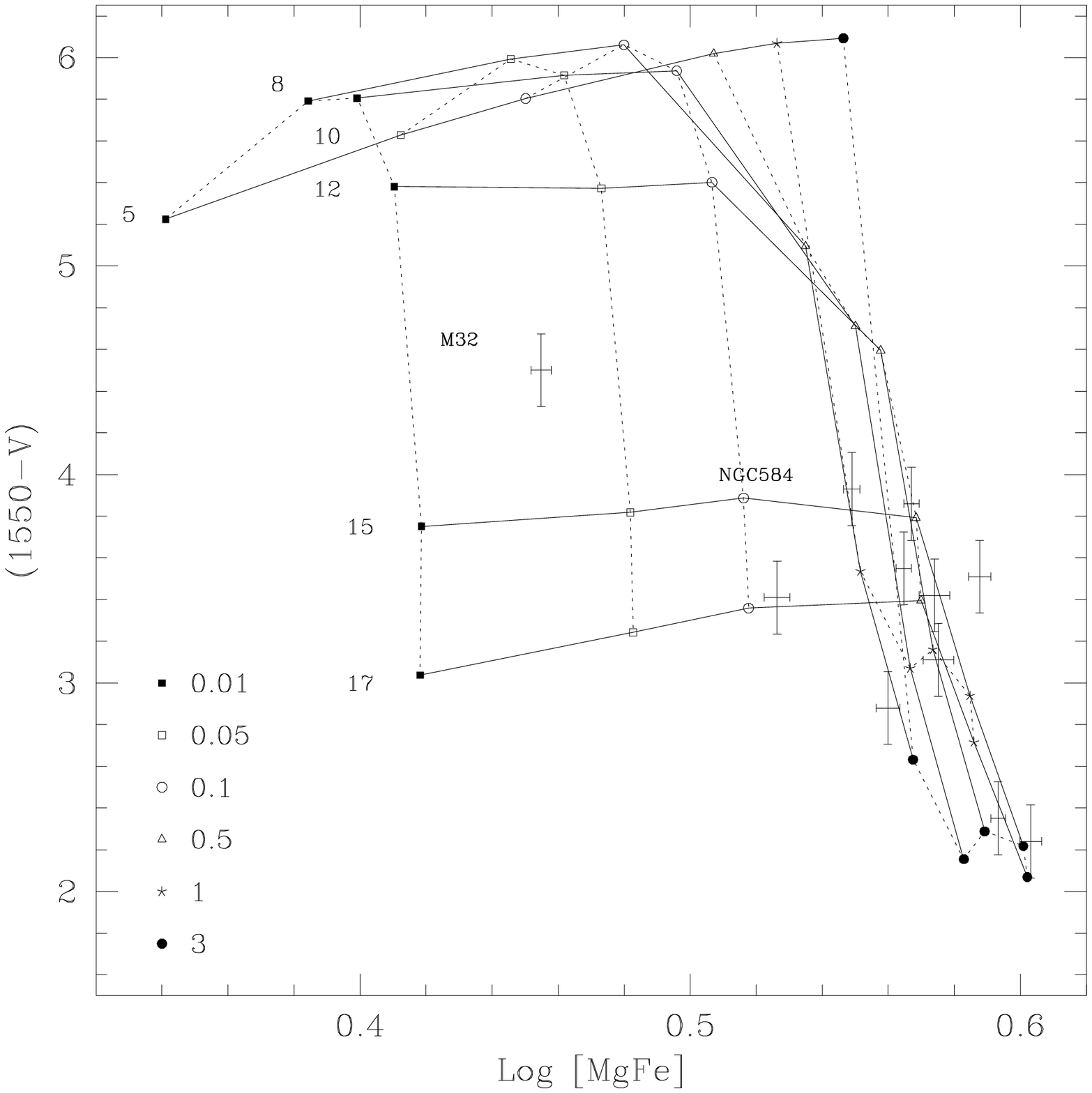,height=9.5truecm,width=8.5truecm}
\caption{ The relation between \UVex\ and \MgFe. The indices \MgFe\  are from
Gonzales (1993), whereas the colours \UVex\ are from Burstein et al. (1988).
The various symbols have the same meaning as in Fig.~\protect\ref{uv_hb}  } 
\label{uv_mg}
\end{figure}

\section{The four-dimensional space \Hbeta\ - \MgFe\ - \UVex\ and $\Sigma$ }

The narrow
 range of the mean and
maximum metallicities inferred from the 
 \Hbeta\ - \MgFe\ plane (with the few
exceptions to be discussed separately), 
 their relatively high values suggested by  the
comparison with theoretical models (see also the summary on observational
hints), and the provisional hypothesis that all galaxies are old nearly
coeval objects 
would imply that  in contrast
to the observations (see Table~\ref{tab_data}) all galaxies of our sample
should emit UV radiation at comparable levels (see the variation 
of the \UVex\ colour as a function of the age for the 0.5 and 3 \ML\ 
galaxies).  Unless galaxies are younger than a few Gyr, this remark would 
apply also to the case in which a significant age difference from galaxy to 
galaxy is allowed.

The situation is  shown in the  Fig.~\ref{uv_hb} and ~\ref{uv_mg} displaying
the separated relations \Hbeta\ - \UVex\  and  \MgFe\ - \UVex\ for the galaxies
in Table~\ref{tab_data}. As well known the \UVex\ colour gets bluer (stronger
UV flux)  with increasing \MgFe\ and decreasing   \Hbeta. The data are compared
with the locus of theoretical models with different \ML, and mean metallicity
in turn, and different age. The solid and dashed lines in Figs.~\ref{uv_hb} and
~\ref{uv_mg} show the loci of constant mass and constant age, respectively. 
The comparison
is made using the Re/8-data because the \UVex\ colours refer to the central
region of the galaxies (cf. Burstein et al. 1988).

Our model galaxies with \dydz=2.5 emit UV radiation by old H-HB stars only for
ages older than about 5.6 Gyr and in presence of suitable metallicities.
Excluding the very initial stages with ongoing star formation in which strong
UV flux can be generated, the colour \UVex\ quickly increases to a maximum and
then decreases again (cf. Bressan et al. 1994, Tantalo et al. 1995). This means
that in the age range 1 to 6 Gyr the \UVex\ - \Hbeta\ relation is almost
insensitive to the age and very close to the 5 Gyr isochrone displayed in
Fig.~\ref{uv_hb}. Similar considerations hold for the diagram \UVex\ - \MgFe\
of Fig.~\ref{uv_mg}.

In the  \UVex\ - \Hbeta\ plane (Fig.~\ref{uv_hb}), 
but for two objects (M32 and NGC~584), all remaining galaxies 
 have data compatible with  ages from 8 to  about 15 Gyr. 
 The \Hbeta\ index of M32 and
NGC~584  is too blue for their \UVex\ colour. The observed   \UVex\ and 
\Hbeta\ of M32, in particular,
 could be matched only going to ages as young as 
about 1 Gyr.  However, in such a case  the corresponding \BV\ colour
would be much bluer than the observational value of 0.85. 

No such discrepancy is found in the \UVex\ - \MgFe\ plane of Fig.~\ref{uv_mg}
where all galaxies are compatible with old ages from 8 to 17 Gyr.
 M32 in particular seems to have
an age of about 13 Gyr.

\begin{figure}
\psfig{file=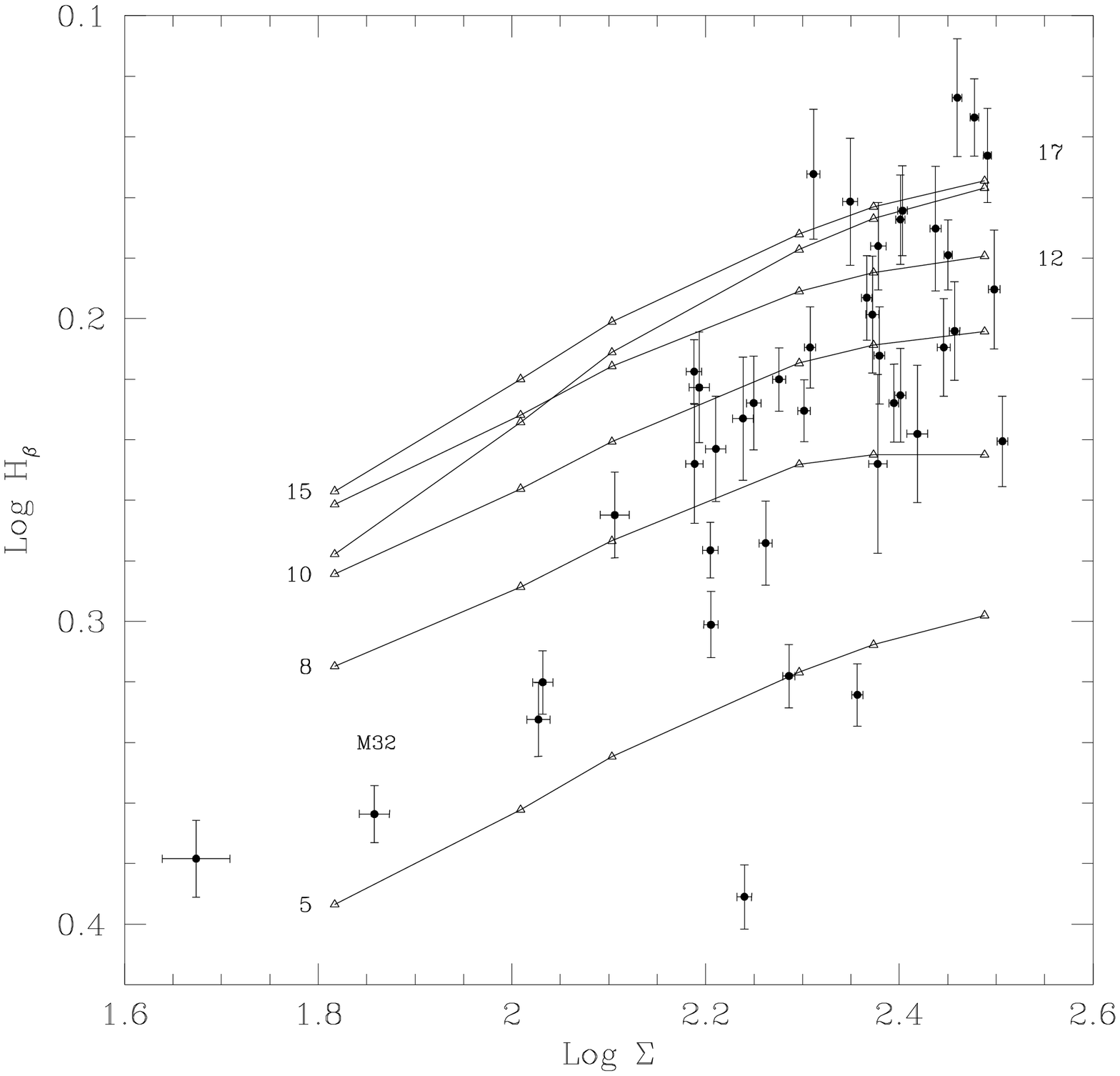,height=8.5truecm,width=8.5truecm}
\caption{ The relation between \Hbeta\ and velocity dispersion  $\log\Sigma$
(in Km sec$^{-1}$). The crosses show the uncertainty affecting the 
observational data. Lines of constant age (in Gyr)  are shown. Finally, the
position of M32 is put into evidence } 
\label{hb_sigma}
\end{figure}

\begin{figure}
\psfig{file=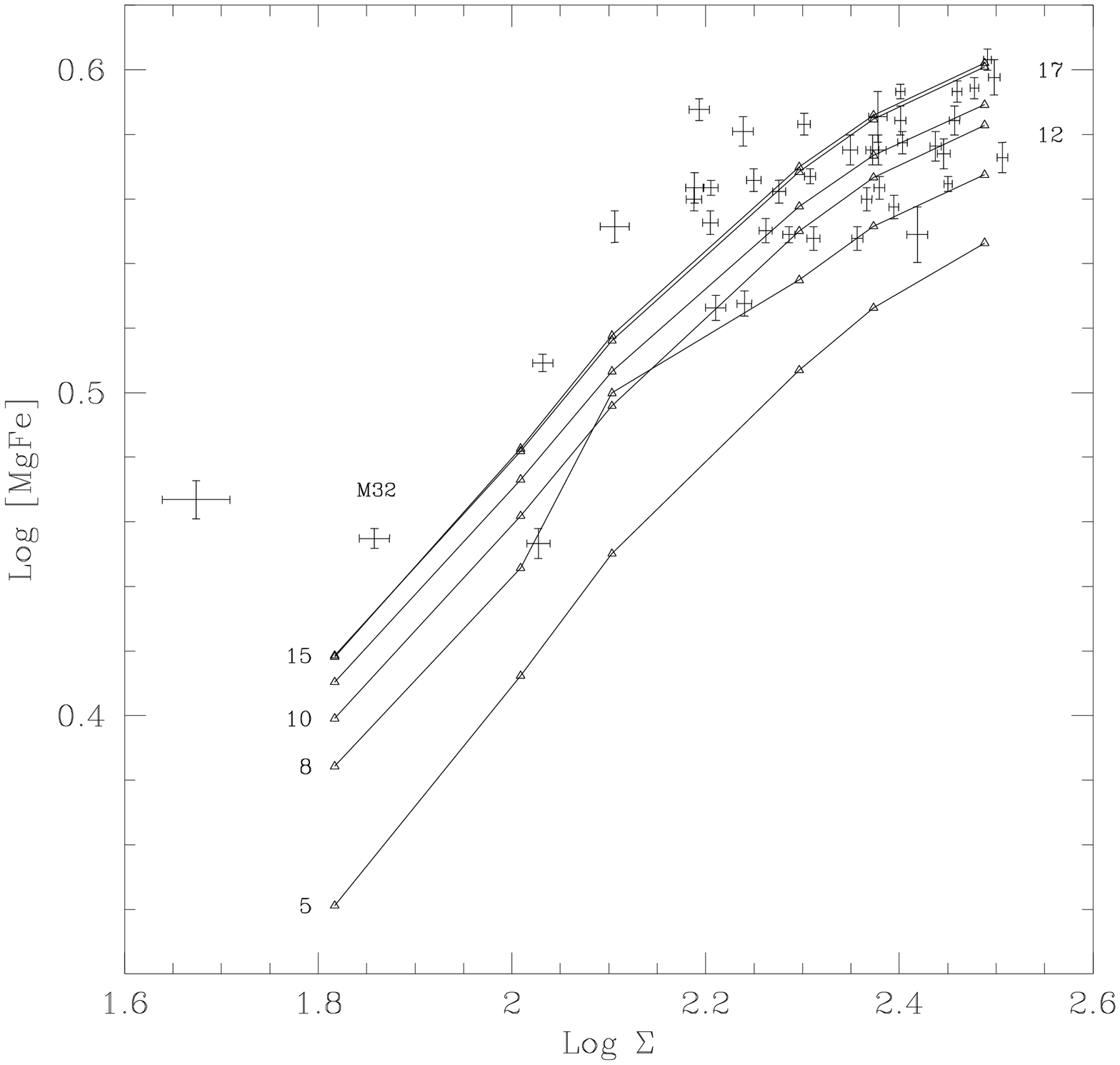,height=8.5truecm,width=8.5truecm}
\caption{ The relation between \MgFe\ and velocity dispersion  $\log\Sigma$ (in
Km sec$^{-1}$). The meaning of the symbols is as in 
Fig.~\protect\ref{hb_sigma} } 
\label{mg_sigma}
\end{figure}

\begin{figure}
\psfig{file=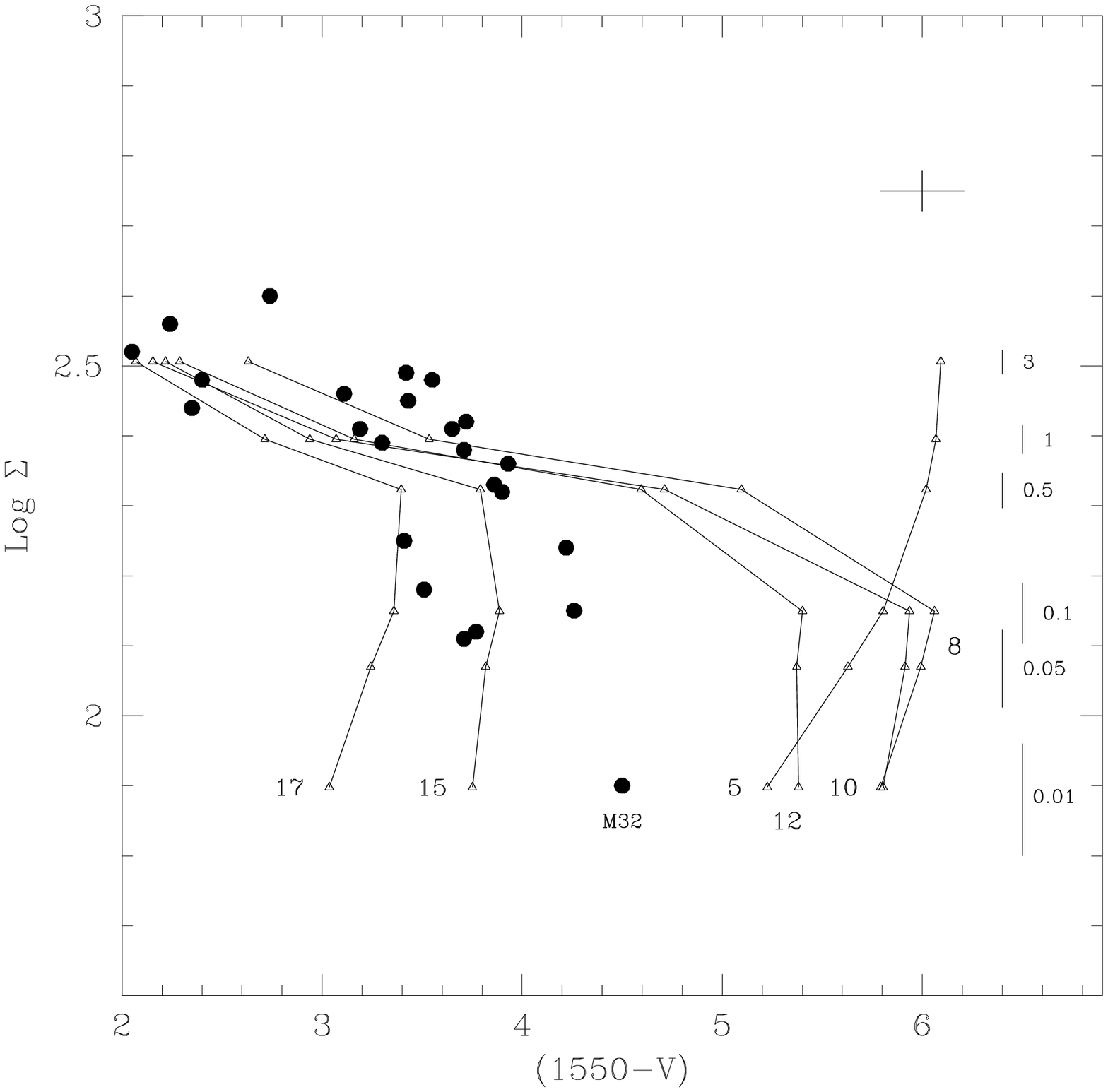,height=8.5truecm,width=8.5truecm}
\caption{ The relation between \UVex\ and velocity dispersion $\log\Sigma$ (in
Km sec$^{-1}$). Lines of constant age (in Gyr) are shown. The vertical bars
indicate the uncertainty in the theoretical $\log\Sigma$ for the galaxies with
different \ML. The cross show the uncertainty of the observational data.
Finally, the position of M32 is shown   } 
\label{uv_sigma}
\end{figure}

To improve upon the above 
analysis  we make use of three additional
observational hints, i.e. the \Hbeta\ - \logS, \MgFe\ - \logS,  and \UVex\ -
\logS\ relations. The data  refer to the central regions of the galaxies (the
Re/8-data set of Table~\ref{tab_data}).
For the sake of internal homogeneity, we adopt 
the \UVex\ and \logS\ data of Burstein et al. (1988) limited to those objects
classified by the authors as quiescent (no evidence of active star formation). 

In order to estimate the velocity dispersion of galactic models we
made use of the  gravitational
potential $\Omega$ and a rough estimate of the radius of the luminous
component. The gravitational potential  is taken from 
Bertin et al. (1992) and
Saglia et al. (1992),

\begin{equation}
\Omega = -  G ({\ML^2 \over \RL}) (\alpha + {\MD \over \ML } \Omega_{LD}')
\end{equation}
where $G$ is the gravitational constant and 

\begin{equation}
\Omega_{LD}' = ({1 \over {2\pi}}) ({\RL \over \RD}  [1+ 1.37 {\RL \over 
\RD}])
\end{equation}
For $\alpha=0.5$,  \ML/\MD=0.2 and \RL/\RD=0.2 as in  Bertin et al. (1992) and
Saglia et al. (1992 ) we get 

\begin{equation}
\Omega= - 0.7 G {M_L^2 \over \RL }
\end{equation}
The radius \RL\ is derived from the
Saito (1979a,b) and Arimoto \& Yoshii (1986, 1987) relationships
 
\begin{equation}
\RL\ = 26.1\times M_L ^{0.55}   ~~~~{\rm kpc}
\end{equation}
Finally, the velocity dispersion $\Sigma$ as a function of \ML\ is 
\begin{equation}
\Sigma = 260.61 \times {\ML }^ {0.225} ~~~~{\rm km~sec^{-1} }
\end{equation}

It goes without saying that the theoretical velocity dispersion is 
affected by a large degree of uncertainty, a rough estimate of which 
can be  evaluated considering \ML\ as a  sort of parameter ranging from the 
original value down to \MS\ and taking for $\Sigma$ the mean of the 
two estimates
($\Sigma_{ad}$). All these quantities are given in Table~\ref{tab_sigma}. 

In Fig.~\ref{hb_sigma} we compare the theoretical 
 \Hbeta\ - \logS\ relations   for different values of the
age with the  data. The scatter both
in \Hbeta\ and \logS\ is large, with galaxies falling along isochrones and
lines of constant mass  spanning an ample range of values. However, the hint
arises that  galaxies with large velocity dispersion are on the average older
than galaxies with low velocity dispersion. In this diagram, M32 has an age of
about 6-7 Gyr. 

In Fig.~\ref{mg_sigma}
we compare the theoretical  \MgFe\ - \logS\ relations with the data.
  But for a general agreement between the slopes of the
observational and theoretical relation, the scatter of the data is so large
that firm indications about the underlying ages are not possible. For the
particular case of M32 an old age is most likely. There are galaxies (NGC~4489
and NGC~7562) that apparently deviate from the mean trend. 

Finally, the theoretical \UVex\ - \logS\ relations are superposed to the data
in  Fig.~\ref{uv_sigma}. Considering all the uncertainties, we draw the
provisional conclusion that all objects of this sample are old. However, 
looking at 
the particular case of M32, it could be either
 as old as about 13 Gyr or as young as about 5 Gyr.

What we learn from the above analysis is that galaxies tend to cluster into
two groups. In the first one (NGC~4649 as prototype) galaxies have
old ages (say from 8 to 17 Gyr) and  
normal behaviour in the various planes. We cannot say whether the large age
range is real or caused by the many uncertainties affecting the analysis.
In the second  group (M32 as a prototype), galaxies 
alternatively change from old to young according to the relationship 
under consideration, which is an obvious  point of contradiction.

\begin{figure}
\psfig{file=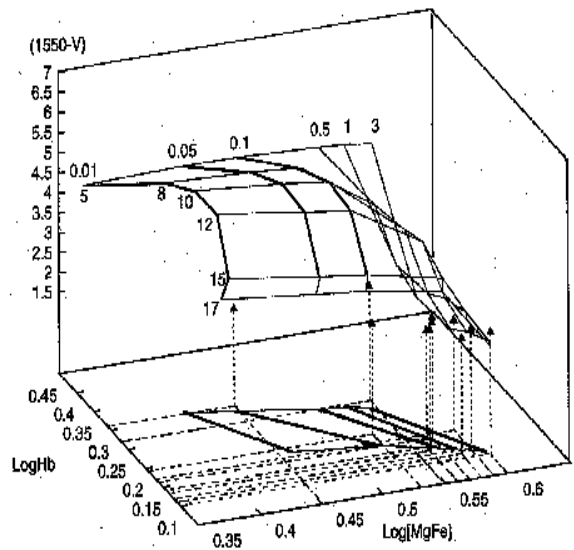,height=8.5truecm,width=8.5truecm}
\caption{The three dimensional correlation between \Hbeta, \MgFe\ and \UVex\
using the Re/8-data of Gonzales (1993) and the UV-fluxes  of Burstein et al.
(1988). The data are shown by the vertical arrows. The thick lines are the loci
of constant mass (in units of \ML), whereas  the thin lines are the loci of
constant age (in Gyr) of the galactic models. These loci are also projected
onto the \Hbeta\ - \MgFe\ plane for the sake of clarity. See the text for more
details  } 
\label{hb_mg_uv}
\end{figure}

To explore further the cause of the above discrepancy,  we look at
the three-dimensional \Hbeta\ - \MgFe\ - \UVex\ relation  for the small group
of galaxies for which the three parameters are simultaneously 
known (11 objects in total,
cf. Table~\ref{tab_data}), and 
 and compare it with the
galactic models of 0.5 and 3 \M12. The \Hbeta\ - \MgFe\ - \UVex\ relation is
shown in Fig.~\ref{hb_mg_uv} for the Re/8-data. 
 The various lines connect loci of constant age (thin)
and constant mass (thick). 

The inspection of  the data in this space reveals that there is a group of
galaxies (NGC~4649 as a prototype) whose age is confined in the range 13 $\div$
15 Gyr (no better estimate is possible) and whose colour \UVex\ is normal and
compatible with the theoretical expectation for old objects containing a
certain fraction of high metallicity stars. Differences from object to object
can be accounted for by differences in mass and hence mean metallicity and
perhaps in  age by as much as  3 Gyr (in agreement with the estimate from the
colour dispersion $\sigma_{\rm (U-V)}$). 

There is another group of galaxies (M32 and NGC~584 as  prototypes) whose
\Hbeta\ is too blue for their \MgFe\ and  \UVex. It seems as if in the \Hbeta\
- \MgFe\ plane an old object has been shifted along the line pertinent to its
mass  by some recent episode of star formation which has changed \Hbeta\
without  changing  \MgFe\ and  \UVex\ significantly. The episode should not
have occurred more recently than about 1  Gyr ago, otherwise the \UVex\ colour
would have changed.

Among the various hypotheses that could be invoked to explain the puzzling
distribution of data in the space of the parameters \Hbeta,  \MgFe,  \UVex, and
\logS,  namely very low metallicities, large range of absolute ages, and burst
of star formation, this latter seems to be the only viable solution. 
 
We have already shown in Fig.~\ref{burst_mod} and Fig.~\ref{hb_burst} how in
the \Hbeta\ -  \MgFe\ plane an old, red galaxy would be rejuvenated by the
occurrence of a burst of stellar activity. The extension and duration of the
loops in the \Hbeta\ -  \MgFe\ plane depends on the intensity and duration of
the burst, and whether or not chemical enrichment takes place. Since the
detailed exploration of space of parameters characterizing a burst is beyond
the scope of this study (in any case no independent observational hints are
available), we limit ourselves to a few general considerations. 

In the rather idealized case that a burst of star formation can occur with no
accompanying chemical enrichment, the net effect would be a narrow loop along
the evolutionary path of a galaxy of given mass. The recovery time scales for
\Hbeta\ and colour \UVex\ are slightly different: the former depending on the
intensity of the burst can be as long as 1$\div$ 1.5 Gyr, whereas the latter as
soon as star formation is over quickly goes back to the previous values.   As
chemical enrichment is excluded, no effects (or very small) can be seen in the
\MgFe\ index.

\begin{figure}
\psfig{file=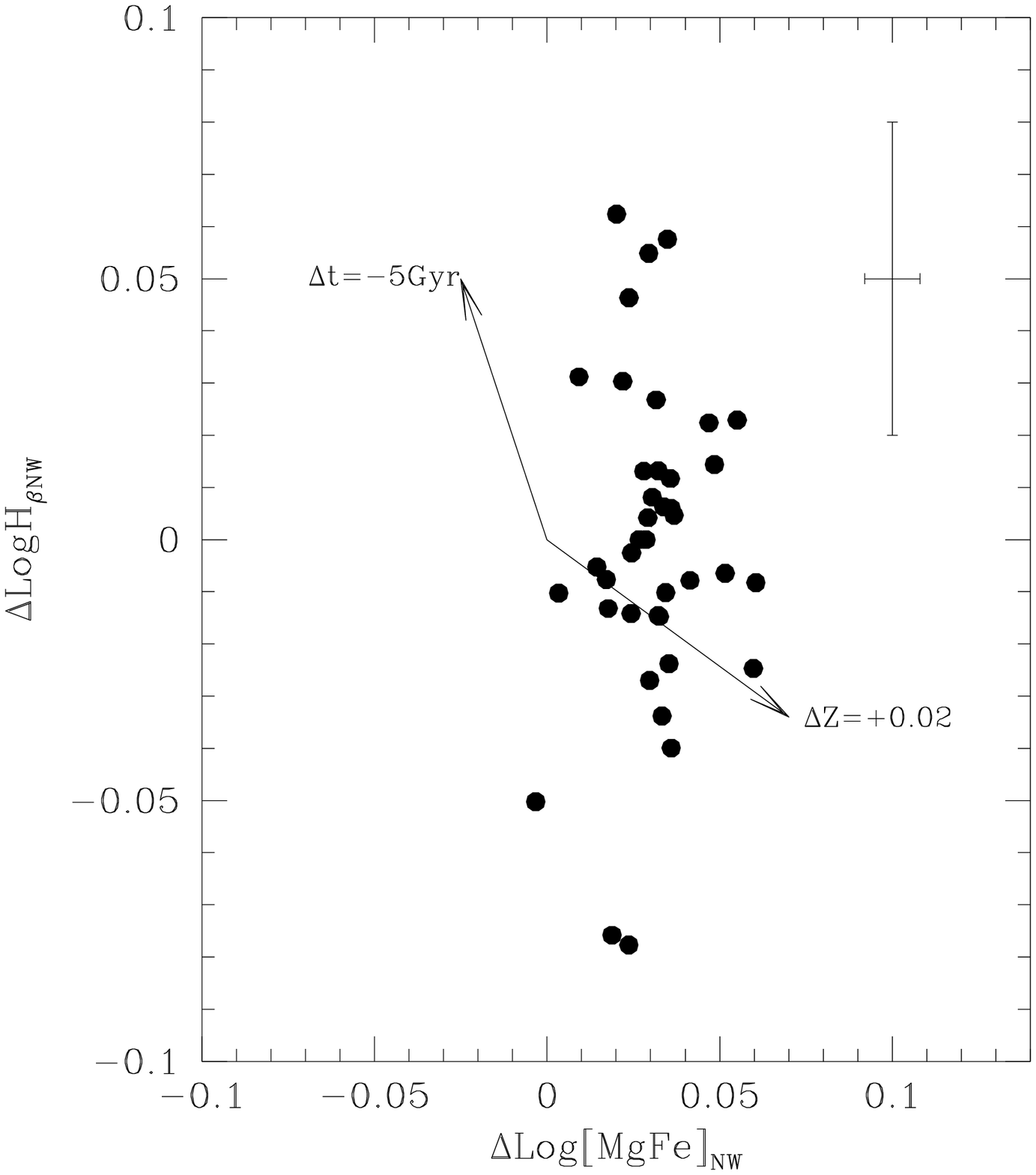,height=8.5truecm,width=8.5truecm}
\caption{ The $\Delta\log\Hbeta_{NW}$ versus $\Delta\log\MgFe_{NW}$ relation.
The two arrows  centered on (0,0) indicate the ageing and enriching vectors of
galaxies and fix the system of coordinates used  to convert $\Delta\Hbeta_{NW}$
and  $\Delta\MgFe_{NW}$ into  $\Delta t_{NW}$ and  $\Delta Z_{NW}$,
respectively. The cross shows the estimated uncertainty of the data.
See the text for more details } 
\label{polar}
\end{figure}

\begin{figure}
\psfig{file=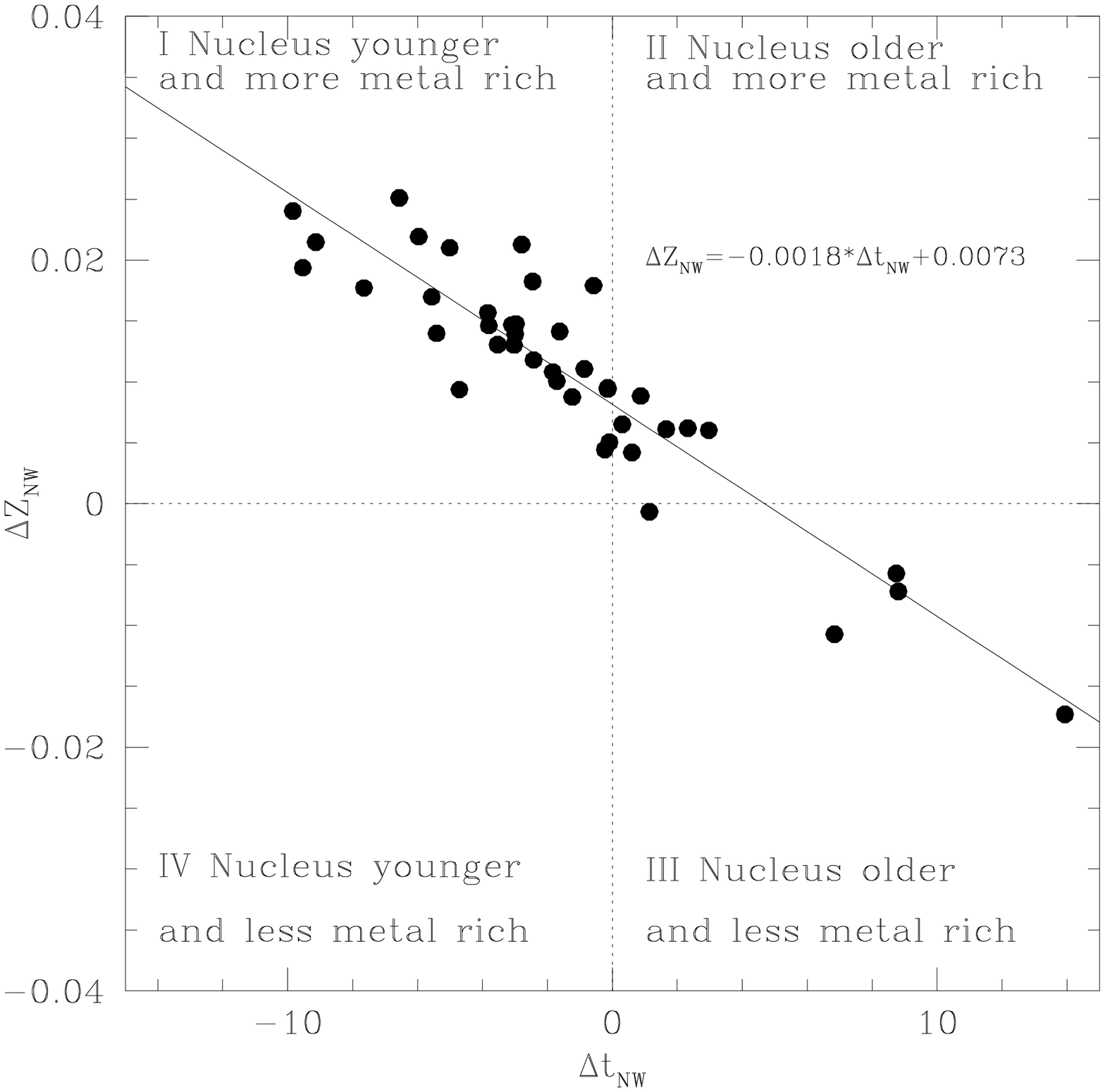,height=8.5truecm,width=8.5truecm}
\caption{The $\Delta t_{NW}$ versus $\Delta Z_{NW}$ relation. The meaning of
the four quadrant's is discussed in the text. The solid line is the linear
best-fit of the data} 
\label{new_sys}
\end{figure}

In the more realistic case, in which chemical enrichment accompanies the star
burst, the evolution in the \Hbeta\ -  \MgFe\ plane is more complicated,
because both \Hbeta\ and   \MgFe\ are now changed by the burst. The increase in
metallicity would shift the galaxy (or part of it) to redder values  of \MgFe\
while performing the loop in the \Hbeta\ -  \MgFe\ plane. 

The hypothesis of a burst of star formation while offering an elegant way out to
the difficulties encountered above, renders the interpretation of these data
more uncertain and arbitrary. 

The key difficulty with this hypothesis, is that in the case of a primordial
origin of elliptical galaxies (all coeval within the age range indicated by the
CMR), a mechanism synchronizing the bursting activity among galaxies would be
required.

In the other alternative that all or a number of elliptical galaxies are the
product of a merger, the star burst hypothesis is viable and can account for the
scatter along the \Hbeta\ axis. 
The only marginal requirement to be satisfied is that the bursting activity
does not significantly change the global pre-burst metallicity 
distribution (see the narrow range of metallicities implied by the 
distribution in the \Hbeta\ - \MgFe\ plane). This is 
 the case in which most of the stars in the 
merging galaxies are formed at an early epoch and some small amount of
star formation occurs at the merger epoch.   
The opposite alternative, in which most of the stars are formed in 
the burst,  while easily explaining the scatter in \Hbeta, could perhaps 
introduce a scatter in metallicity much larger than that inferred from the 
observations.

\section{ Age and metallicity gradients in galaxies}

The galaxies in our sample possess gradients in \Hbeta\ and \MgFe\ passing from
the Re/2 to the Re/8-data set (cf. Gonzales 1993 and the entries of
Table~\ref{tab_data}) that  likely reflect gradients in age and/or
 metallicity. In
order to clarify the role played by these gradients in understanding the
structure and the evolution of galaxies, we perform the following analysis. Let
us define the  quantities 

\begin{equation}
  \Delta\log\Hbeta_{NW} = \log\Hbeta_N - \log\Hbeta_W
\label{beta_NW}
\end{equation}

\begin{equation}
  \Delta\log\MgFe_{NW} = \log\MgFe_N - \log\MgFe_W
\label{MgFe_NW}
\end{equation}
where  N and W  stand for the  Re/8- and R/2-data set, i.e. for the central
regions and the whole  galaxy, respectively. 

The relation between $\Delta\log\Hbeta_{NW}$ and  $\Delta\log\MgFe_{NW}$ is
shown in Fig.~\ref{polar}, in which the cross shows an estimate of the 
typical uncertainty affecting the data. The size of the error bars is 
assumed to be twice   the mean value of the errors in the data 
of Fig.~\ref{gonz_data}.

Despite the above uncertainty, 
with the aid of equations (13) through (16) we draw
in Fig.~\ref{polar} the system of coordinates represented by the two vectors
$\Delta Z=0.02$ and $\Delta t= -5$ Gyr  and centered on the (0,0) point of the
$\Delta\log\Hbeta_{NW}$ - $\Delta\log\MgFe_{NW}$ plane. The new system of
coordinates represents the ageing and enriching vectors of galaxies or parts of
these. Projecting the data of the  $\Delta\log\Hbeta_{NW}$ -
$\Delta\log\MgFe_{NW}$ plane onto the new system of coordinates we get the mean
age and metallicity differences, $\Delta t_{NW}$ and  $\Delta Z_{NW}$,
respectively, between the central regions and the whole galaxy. These
quantities are given in Table~\ref{zeta_age} and their correlation is plotted
in Fig.~\ref{new_sys}.

\begin{table*}
\caption{Data on the $\Delta\log\Hbeta_{NW}$ - $\Delta\log\MgFe_{NW}$ plane and
mean metallicity and age differences between the nucleus and the whole galaxy,
$\Delta Z_{NW}$ and    $\Delta t_{NW}$, respectively.   } 
\littleskip
\scriptsize
\begin{center}
\begin{tabular*}{170mm}{c c c c c c c c c c c   } 
\hline 
\hline
NGC  &   $\Delta\log\MgFe_{NW}$&$\Delta\log\Hbeta_{NW}$
                               &$\Delta Z_{NW}$ & $\Delta t_{NW}$ &  &
NGC  &  $\Delta\log\MgFe_{NW}$&$\Delta\log\Hbeta_{NW}$
                               &$\Delta Z_{NW}$ & $\Delta t_{NW}$\\   
\hline
     &       &       &       &       & &    &       &       &        &       \\
\hline
  221&  +0.00&  +0.03& +0.006& -3.962& &4472&  +0.01&  -0.01& +0.002 & +0.679\\ 
  224&  +0.01&  -0.01& +0.002& +0.679& &4478&  +0.03&  +0.02& +0.015 & -4.566\\
  315&  +0.03&  -0.01& +0.009& -0.604& &4489&  +0.05&  +0.02& +0.023 & -5.849\\
  507&  +0.02&  -0.07& -0.006& +7.962& &4552&  +0.03&  -0.01& +0.009 & -0.604\\
  547&  +0.03&  +0.05& +0.021& -8.528& &4649&  +0.03&  +0.01& +0.013 & -3.245\\
  584&  +0.03&  +0.01& +0.013& -3.245& &4697&  +0.06&  +0.02& +0.026 & -6.491\\
  636&  +0.03&  +0.01& +0.013& -3.245& &5638&  +0.04&  -0.01& +0.013 & -1.245\\
  720&  +0.01&  -0.11& -0.017& 13.887& &5812&  +0.02&  +0.00& +0.008 & -1.283\\
  821&  +0.03&  -0.04& +0.004& +3.358& &5813&  +0.02&  +0.06& +0.019 & -9.208\\
 1453&  +0.03&  -0.03& +0.006& +2.038& &5831&  -0.01&  -0.01& -0.006 & +1.962\\
 1600&  +0.00&  -0.05& -0.009& +6.604& &5846&  +0.04&  +0.06& +0.026 & -10.491\\
 1700&  +0.03&  +0.00& +0.011& -1.925& &6127&  +0.03&  +0.00& +0.011 & -1.925\\
 2300&  +0.02&  +0.02& +0.011& -3.925& &6702&  +0.02&  -0.01& +0.006 & +0.038\\
 2778&  +0.03&  +0.06& +0.023& -9.849& &6703&  +0.04&  +0.01& +0.017 & -3.887\\
 3377&  +0.06&  -0.01& +0.021& -2.528& &7052&  +0.03&  -0.08& -0.004 & +8.642\\
 3379&  +0.03&  +0.01& +0.013& -3.245& &7454&  +0.05&  +0.01& +0.021 & -4.528\\
 3608&  +0.04&  -0.01& +0.013& -1.245& &7562&  -0.03&  -0.01& -0.013 & +3.245\\
 3818&  +0.06&  -0.03& +0.017& +0.113& &7619&  +0.03&  -0.04& +0.004 & +3.358\\
 4261&  +0.03&  +0.02& +0.015& -4.566& &7626&  +0.04&  +0.00& +0.015 & -2.566\\
 4278&  +0.03&  -0.03& +0.006& +2.038& &7785&  +0.02&  +0.03& +0.013 & -5.245\\
 4374&  +0.02&  -0.01& +0.006& +0.038& &    &       &       &        &  \\
\hline 
\hline
\end{tabular*}
\end{center}
\label{zeta_age}
\normalsize
\end{table*}

The plane of Fig.~\ref{new_sys} is divided in four quadrant's  characterized by
the sign (either positive or negative) of $\Delta t_{NW}$ and $\Delta Z_{NW}$.
The meaning of the four quadrant's is as follows: 
\begin{itemize}

\item{ $\Delta t_{NW} < 0$ and $\Delta Z_{NW} > 0$: the nucleus is younger and
more metal-rich than the external regions of the galaxy. This corresponds to a
sort of out-inward  process of galaxy formation, in which star formation in the
nucleus continued for significant periods of time, up 10 Gyr in some cases.
Looking at number of galaxies in this regions, this case seems to be the most
probable in nature. } 

\item{$\Delta t_{NW} > 0$ and $\Delta Z_{NW} > 0$: the nucleus is expected to
be older and more metal-rich than the external regions. Only a few galaxies are
found in this quadrant but with  small differences in the age and metallicity.
These galaxies could have been formed by a sort of in-outward mechanism  on a
rather short time scale.  } 

\item{$\Delta t_{NW} > 0$ and $\Delta Z_{NW} < 0$: the nucleus is older but
less metal-rich than the external regions.  These objects do not find a
straightforward explanation. They  could correspond to cases of merge in which
an old compact primeval object with scarce chemical enrichment subsequently
captured other smaller galaxies in which chemical enrichment has already
proceeded to
substantial levels. } 

\item{ $\Delta t_{NW} < 0$ and $\Delta Z_{NW} < 0$: the nucleus should be
younger and less metal-rich than  the external regions. No galaxy is found in
this quadrant. This kind of galaxy structure and star formation in turn is not
allowed. } 

\item{ There appears to be a tight correlation between $\Delta Z_{NW}$ and
$\Delta t_{NW}$ as indicated by  the linear best-fit 
\begin{equation}
\Delta Z_{NW} = -0.0018 \times  \Delta t_{NW} + 0.0073 
\end{equation}
with $\Delta t_{NW}$ in Gyr. This relation indicates that galaxies as a whole
or their constituent parts in spite of the large variety of properties (masses,
dimensions, colours, luminosities, etc.) are ultimately  governed by age and
metallicity, and that chemical enrichment in these obeys an universal law. } 
\end{itemize}

This analysis cannot  yield the true ages and metallicities of the nuclear and
peripheral regions of a galaxy but only their mean difference. 

It goes without saying that the above results depend on the transformation from
the $\Delta\log\Hbeta_{NW}$ - $\Delta\log\MgFe_{NW}$ to the $\Delta t_{NW}$ -
$\Delta Z_{NW}$ plane. However, the transformation in use cannot be grossly in
error because it stems from the properties of SSPs and model galaxies. As
already amply  discussed in the previous sections both match the data in the
basic $\Hbeta$ - $\MgFe$. 

The numerous group of galaxies in the first quadrant ($\Delta t_{NW} < 0$ and
$\Delta Z_{NW} > 0$) is of particular interest. In spite of the afore 
mentioned uncertainty, for these galaxies we suggest
that a sort of out-inward process of formation ought to have occurred, lasting
in many  cases up to several Gyr.  In this context,  it would be interesting to
know whether the time $\Delta t_{NW}$ correlates with other important
parameters such as the velocity, dispersion \logS, the mass to blue luminosity
ratio $(M/L_B)_{\odot}$, the effective radius Re, and the total mass. The
inspection of the data contained in Tables~\ref{tab_data} and ~\ref{zeta_age},
indicates
that there could be a weak correlation between $\Delta t_{NW}$ and the velocity
dispersion $\Sigma$: 
\begin{equation}
\Delta t_{NW} = 3.327\times \logS\ - 10.65 
\end{equation}
where $\Delta t_{NW}$ is in Gyr, $\Sigma$ is in ${\rm km~sec^{-1}}$, and the
correlation coefficient is 0.43. This relation has been derived discarding  the
four objects with the longest, perhaps suspected,  $\Delta t_{NW}$, namely
NGC~547, NGC~2778, NGC~5813 and NGC~5846. Similar but weaker dependencies on
the total mass and effective radius can be found. 

Therefore, the hint arises  that  the time difference between the global and
the nuclear activity  decreases at increasing velocity dispersion and perhaps
total mass. This time difference can be taken as an indication of the overall
duration of the star forming period.

\section{Concluding remarks}

In this paper we have addressed the question whether age and metallicity 
effects on the stellar content of elliptical galaxies 
can be disentangled by means of broad-band colors and line strength 
indices derived from detailed chemo-spectro-photometric models. The main 
goal of our study is to cast light on a number of interwoven questions:
are  elliptical galaxies made only of old stars generated in a dominant 
initial episode of star formation? is there any evidence of subsequent star 
forming  episodes ? If so, has this activity an internal
origin or is it induced by the merger mechanism ? 

The analysis of the observational data in the space of
the parameters \Hbeta, \MgFe, \logS, and \UVex\ and of the gradients in\Hbeta\
and \MgFe\ within individual galaxies allow us to draw a number of conclusions.

\begin{itemize}
\item{The  distribution of galaxies in the \Hbeta\ - \MgFe\ plane does  not 
correspond to
 a mere  sequence of metallicity with bluer galaxies significantly 
more metal-poor
 than the red ones. Metallicity effects are expected to broaden the
distribution mainly along the   \MgFe\ axis.} 

\item{Equally, the  distribution of galaxies in the  \Hbeta\ - \MgFe\ plane 
does not constitute 
 a mere sequence of age with  bluer galaxies significantly younger than the
red ones. Although some scatter in the age is possible, it is not likely to be
as large as indicated by the formal match of data with isochrones in this
diagram. M32 in particular according to its \Hbeta\ should have an age of about
3 Gyr, in disagreement with its \UVex=4.5 for which much older ages are needed
according to the present theory of population synthesis.} 

\item{The observed distribution of galaxies in the  \Hbeta\ - \MgFe\ plane does
not agree with the CMR-strip on the notion that they evolve as passive objects
in which star formation took place in an initial episode. Most likely, the
history of star formation was more complicated than this simple scheme. } 

\item{The observed \UVex\ colours are not compatible with ages younger that
about 6 Gyr. UV excesses stronger than \UVex=5.5 at younger ages are possible
only if the age is younger that $1\div 2$ Gyr. However, in such a case the
remaining UBV colours would be too blue compared with the observational data.
Most likely, all galaxies in the sample are globally 
older than at least 6 Gyr. } 

\item{As far as ages are concerned, our analysis is still unable to give a 
definitive answer. Besides the group of galaxies like M32 and NGC~584 whose 
properties lead to contrasting results depending on the particular 
relationship among the four parameters \Hbeta, \MgFe, \UVex\ and $\Sigma$
under examination, the remaining galaxies appear to span a wide range of 
ages going from  8 to 15 Gyr. However, this large scatter can be partly 
due to observational uncertainties and partly  to a real age spread. 
It is worth pointing out that for the very small subgroup of galaxies with  
all the four parameters simultaneously  available, the objects with an old age
($13\div 15$ Gyr) like NGC~4649 show an  age spread  amounting only  to 
about 3 Gyr, which is   consistent with the small scatter in the
CMR of cluster galaxies.}

\item{The present analysis cannot yet cast light on 
 the dominant mechanism by which  elliptical galaxies are formed. However, 
the following considerations can be made.
 If elliptical galaxies are all old, coeval  and evolving in isolation, the
difficulties encountered with the interpretation of the \Hbeta, \MgFe, \UVex,
and \logS\ data can be removed by invoking later mild 
episodes of star formation. 
 The major problem with this suggestion  is that in the case of 
evolution in isolation, 
some mechanism synchronizing the bursts of star formation should be required in
order to dislocate galaxies from the CMR-strip to their observed location in
the \Hbeta\ - \MgFe\ plane. Indeed, even if the statistics is limited, no
galaxies are seen along the CMR-strip but for those at the red end.  
A simple way out could be that the star forming activity is limited to the
nuclear region, thus affecting the \Hbeta\  index without changing 
significantly the global value of the broad-band colours (cf. the case of 
NGC~4374 and NGC~4697 for which little age scatter is indicated by the 
CMR, whereas a significant age scatter is suggested by the indices). 
No such difficulty would exist if already formed ellipticals 
suffer in more recent times from  
interaction with other small size galaxies (for instance the capture of 
small gas rich systems) thus inducing some mild star formation activity
with some chemical enrichment.
Equally, if  elliptical galaxies are the result of classical mergers
of spiral galaxies. In such a case a spread in \Hbeta mimicking an 
 age sequence for the resulting composite 
population would naturally follow. The accompanying chemical 
enrichment can be easily adjusted to  match the small observational scatter 
along the \MgFe\ axis. }

\item{Although based on a handful of objects, there is a interesting scenario
emerging from the analysis of the four dimensional space \Hbeta,  \MgFe, \UVex,
and \logS. In brief, galaxies of high velocity dispersion 
tend to be confined at the red end of the \Hbeta\ - \MgFe\ distribution.  Their
properties are consistent with being old objects (some scatter in the age is
perhaps possible) that are able to exhaust the star forming process at very
early epochs with no need for later episodes of stellar activity. In contrast,
galaxies of lower velocity dispersion have more contrasting
properties. It seems as if in these systems star formation, following the
initial activity most likely at epochs as old as in other galaxies, later went
through a series of episodes taking place in different epochs that vary from
galaxy to galaxy. Equivalently, it could be that star formation continued
perhaps at minimal levels over significantly longer periods of time. This would
mimic a sort of age  sequence. } 

\item{The analysis of the gradients in \Hbeta\ and \MgFe\ within individual
galaxies has revealed that for the majority of these the nucleus is younger and
more metal-rich than the peripheral regions, thus resembling a out-inward
mechanism of galaxy formation, and that the global process of star formation
may last very long, with duration varying from galaxy to galaxy. It seems that
the duration of the star formation activity is inversely proportional to the
 galactic velocity dispersion and perhaps mass. There are two less numerous
subgroups for which the nucleus is older and more metal-rich or older and
less-metal rich than the periphery of the galaxy. The interpretation of these
structure is not straightforward and it is left to future investigation.} 

\item{The fact that most galaxies are found in the group characterized by a
younger and more metal-rich nucleus calls to mind the fact that recent bursts
of star formation in the merger scenario are expected to peak in the centers of
elliptical galaxies (Alfensleben \& Gerhard 1994). However, an internal cause
related to the building up process of galaxy formation cannot be excluded as
perhaps suggest by the correlation between velocity dispersion and $\Delta
t_{NW}$. } 

 \item{The idea that the overall duration of the star forming activity is
inversely proportional to the velocity dispersion and perhaps  mass of galaxies
does not contradict the current information of abundance ratios in elliptical
galaxies  (Carollo et al. 1993, Carollo \& Danziger 1994, Matteucci 1994).
Indeed it would be consistent with the expected trend for the ratio [Mg/Fe]
inferred from the narrow band indices (Faber et al. 1992; Worthey et al. 1992;
Davies et al. 1993). From the comparison of the observed indices with model
indices, assuming solar partition of elemental abundances, it is concluded that
the average [Mg/Fe] in giant elliptical galaxies exceeds that of the most metal
rich stars in the solar vicinity by about 0.2-0.3 dex. Furthermore, the ratio
[Mg/Fe] is expected to increase with the galactic mass up to the this value.
According to Matteucci (1994) a possible explanation of this trend is that the
efficiency of star formation is an increasing function of galactic mass. As a
consequence of this, massive ellipticals should have formed on shorter time
scales than smaller ellipticals. In brief, the constant super-solar value for
giant ellipticals hints that only an unique source of nucleosynthesis is
contributing to chemical enrichment. This is just the case of massive stars
exploding as type II supernovae and thus releasing elemental species (Mg and Fe
in particular) in fixed proportions. This implies a short time scale of star
formation, i.e. of about 0.1 Gyr. The trend decreasing with galactic mass means
that another source of Fe intervenes altering the [Mg/Fe] ratio. It is worth
recalling that Mg is produced by massive stars. Therefore, it is the
contribution to Fe that must change. The goal is achieved invoking type I
supernovae generated by mass accreting white dwarfs in binary systems. The
minimum time scale for their formation is longer than 0.1 Gyr. This implies a
longer duration of the star forming period. In this context, we would like to
recall that in order to match the CMR with our infall  models we had to
increase the efficiency parameter $\nu$ by a factor of 12 as the galactic mass
increases from 0.01 to 3 \M12. Furthermore, the duration of the star forming
activity determined by the onset of galactic winds was found to decrease with
the galactic mass (see $t_{gw}$ in Table~\ref{tab_mass}). The trend is small
but in the right direction. } 
\end{itemize}

\oneskip

\begin{acknowledgements} 
The authors deeply thank an anonymous referee for so carefully reading 
an early draft of this paper and for making some very useful comments which
improved the manuscript. 
The financial support from 
the Italian Ministry of University, Scientific
Research and Technology (MURST) and the Italian Space Agency (ASI) is also
gratefully acknowledged. 
\end{acknowledgements}

\end{document}